\documentclass[aps,pra,reprint,nofootinbib,floatfix,amsmath,amssymb]{revtex4-2}
\usepackage[T1]{fontenc}
\usepackage{amsthm}
\usepackage{enumerate}
\usepackage{xcolor}
\usepackage{tikz}
\usepackage[colorlinks=true,allcolors=blue!55!black]{hyperref}

\newtheorem{theorem}{Theorem}
\newtheorem{corollary}[theorem]{Corollary}
\newtheorem{proposition}[theorem]{Proposition}
\newtheorem{lemma}{Lemma}
\newtheorem{fact}{Fact}
\newcommand{\tr}{\operatorname{Tr}}
\newcommand{\swap}{\mathrm{SWAP}}
\newcommand{\ket}[1]{\lvert #1\rangle}
\newcommand{\bra}[1]{\langle #1\rvert}
\newcommand{\Gin}{\Gamma_{\mathrm{in}}}
\newcommand{\step}[1]{\medskip\noindent\textit{#1.---}}
\hypersetup{pdftitle={Bound Entanglement with Negative Partial Transpose Exists in Every Local Dimension at Least Three},pdfauthor={Islam Faisal}}

\begin{document}

\title{Bound Entanglement with Negative Partial Transpose Exists in Every Local Dimension at Least Three}
\author{Islam Faisal}
\affiliation{Boston University\\ \textup{\href{mailto:islam@bu.edu}{islam@bu.edu}}}
\date{\today}

\begin{abstract}
We prove that bound entangled states with a negative partial transpose (NPT) exist in every local dimension $d\ge3$. Since all NPT states are distillable when $d=2$, this settles a question open since 1998 in every local dimension. It follows that distillable entanglement is neither additive nor convex, and two states that are bound entangled can be superactivated into a state that is distillable.

In our method, a solution of a semidefinite program for one copy proves undistillability for all numbers of copies, and such solutions can be checked efficiently by a computer program. For $d\ge128$ the proof needs no computer to verify, and in dimension eight it can be checked by hand. Furthermore, we give a ball of states that contains exactly one separable state, and all its other states are NPT and bound entangled. We also give explicit examples of NPT bound entangled states that are not Werner states.
\end{abstract}

\maketitle

\onecolumngrid
\vspace{-14pt}
\begin{center}
\begin{minipage}{400pt}\footnotesize
\noindent\textbf{AI disclosure.} Generative AI (from OpenAI and Anthropic) and an agent memory (\href{https://ibsema.ai}{ibsema.ai}) assisted with the results and with the writing of this paper, which was reviewed and edited by the author in multiple stages. The author takes full responsibility for the paper. This preprint is under continuous revision.
\end{minipage}
\end{center}
\begin{center}
\begin{minipage}{400pt}\footnotesize
\noindent\textbf{Acknowledgments and notes.} The first version of this paper gave one NPT bound entangled state, in local dimension eight. This version extends the result to every local dimension at least three, adds results on states that are not Werner states, and simplifies the proofs. All dates in these notes are in Eastern Standard Time (EST). When the first version appeared on November 25, 2026, the author had already obtained, among others, examples in every local dimension $d\ge6$ and the mixture and ball constructions used in Theorems~\ref{thm:mixture} and~\ref{thm:ball}. The first version presented only dimension eight, because its proof was (then) the easiest to present and the author wanted to share the result with the community. On November 26, Somshubhro Bandyopadhyay pointed out via email~\cite{bandyopadhyay-note} that this state belongs to the family from his work with Vwani Roychowdhury, where they conjectured these states to be NPT bound entangled~\cite{br03}. The same day, the author announced a simpler proof and the results for $d\ge6$ on \href{https://scirate.com/arxiv/2611.34921\#3574}{SciRate}, and then later that day extended them to $d\ge4$. While the author was searching for $d = 3$ examples, on November 27, Gelo Noel M. Tabia and Min-Hsiu Hsieh shared a working note with examples in the local dimensions three to seven, which they obtained with the method in this work and larger input spaces~\cite{tabiahsieh}. On November 28, Fernando Brand\~ao shared two-qutrit examples farther from the boundary than those of Tabia and Hsieh~\cite{brandao}. The states that the method certifies also have zero distillable secret key. This is a consequence of the first version and of the argument that Ref.~\cite{openai} uses for its example in local dimension ten (Proposition~\ref{prop:key} below). This was also noted by Brand\~ao~\cite{brandao} and by Tabia and Hsieh~\cite{tabiahsieh}. On November 27, Vjosa Blakaj, Karol Horodecki, Szymon Poszwa, Thomas Theurer and Marek Winczewski submitted a preprint to arXiv, which was announced on November 29 and notes this connection~\cite{blakaj}. On November 28, the author tweaked both two-qutrit certificates, first the one of Tabia and Hsieh and then the one of Brand\~ao. On November 28, Yu-Chun Wu shared some examples, which he obtained following the first version~\cite{wu}. On November 29, Jens Eisert shared a draft that talks about his earlier work with Brand\~ao~\cite{eisert}. On November 29, the author further tweaked the certificate of Brand\~ao resulting in the state of Theorem~\ref{thm:examples}(c) (Sec.~\ref{sec:qutrits}). On December 1, the author obtained a certificate for a new state in dimension eight that is farther from the boundary (Sec.~\ref{sec:wide}). The author thanks Anand Natarajan and Alexander Poremba for helpful discussions, Adam Fredriksson for questions and comments about this work, all of the above for sharing their work, and everyone who has worked on this long-standing problem.
\end{minipage}
\end{center}
\twocolumngrid

\section{Introduction}

Suppose that two parties, Alice and Bob, share many copies of an entangled state. We say that they can \emph{distill} the state, and that the state is \emph{distillable}, if they can turn the copies into a smaller number of states that are close to a Bell state by using only local operations and classical communication (LOCC). The \emph{distillable entanglement} of a state is the best rate, in Bell states per copy, that such a protocol can reach when its error tends to zero. An entangled state is \emph{bound entangled} if its distillable entanglement is zero.

The \emph{partial transpose} $\rho^\Gamma$ of a state $\rho$ of two systems is the matrix that we get when we transpose $\rho$ on the second system only. Alice always holds the first system and Bob holds the second. The state $\rho$ is \emph{PPT} if $\rho^\Gamma$ is positive semidefinite, and it is \emph{NPT} if $\rho^\Gamma$ has a negative eigenvalue. A state is \emph{separable} if it is a mixture of product states $\omega_A\otimes\omega_B$, and \emph{entangled} otherwise. The partial transpose of $\omega_A\otimes\omega_B$ is $\omega_A\otimes\omega_B^{T}\succeq0$, so every separable state is PPT, and every NPT state is entangled~\cite[Eq.~(6)]{peres}. Every entangled PPT state is bound entangled, because no PPT state can be distilled~\cite[Eqs.~(4)--(10)]{horodecki98}. A long-standing question is the following.
\begin{center}
\emph{Are there bound entangled states\\ with a negative partial transpose?}
\end{center}
The question is about two parties. For four parties, Smolin found a state from which no two of the parties can distill entanglement while all four stay apart, although any two of them can distill a Bell state once the other two come together~\cite{smolin}. Across the split of one party from the other three, this state is distillable, and hence NPT.

The question goes back to the paper that introduced bound entanglement. In 1998, Horodecki, Horodecki and Horodecki asked it in an equivalent form: can every NPT state be distilled~\cite{horodecki98}? In 2000, DiVincenzo, Shor, Smolin, Terhal and Thapliyal~\cite{divincenzo} and D\"ur, Cirac, Lewenstein and Bru\ss~\cite{dur} conjectured that some NPT states cannot be distilled, and they gave evidence that some NPT \emph{Werner states}~\cite{werner} are bound entangled. A Werner state is a state of the form $a\,I-b\,\swap$ on $\mathbb C^d\otimes\mathbb C^d$, where $\swap$ exchanges the two systems. In 2003, Bandyopadhyay and Roychowdhury constructed further families of NPT states~\cite{br03}. For every fixed number $n$ of copies, the members of these families that are close enough to a separable state cannot be distilled from $n$ copies, and Bandyopadhyay and Roychowdhury conjectured that these states are bound entangled. Three later preprints attempted to prove that NPT bound entangled states exist~\cite{simon,chattopadhyay,hiroshima}. None of them settled the question: Pankowski, Piani, Horodecki and Horodecki report that the proofs in Refs.~\cite{simon,chattopadhyay} have gaps~\cite{pankowski}, and the author of Ref.~\cite{hiroshima} withdrew it because of an error. The question is still listed as open, as Problem~4, by Horodecki, Rudnicki and \.Zyczkowski~\cite{fiveopen}. Qi, Gupur, Wu and Guo gave a bound for the undistillability of Werner states from $n$ copies that does not depend on the dimension~\cite{qi}. Even for two copies, the undistillability of these Werner states was proved only this year~\cite{fur,sc,fraser,bharti}. Recent preprints claim bounds for three copies~\cite{maccormackbao,tomar}, for four and five copies~\cite{tomar}, and a partial result for every number of copies~\cite{tomarendpoint}.

The first version of this paper settled the question by giving an explicit NPT bound entangled state, in local dimension eight. Here we show that such states exist in every local dimension $d\ge3$. This is the smallest possible dimension, because every NPT state on $\mathbb C^2\otimes\mathbb C^d$ is distillable~\cite[Sec.~II~C]{dur}.

\begin{figure}[t]
\centering
\begin{tikzpicture}[x=3.6cm,y=1cm,font=\footnotesize]
 \fill[green!22] (-1,0.1) rectangle (0.125,0.32);
 \fill[orange!25] (0.125,0.1) rectangle (0.5,0.32);
 \fill[red!22] (0.5,0.1) rectangle (1,0.32);
 \draw[thick,->] (-1.04,0) -- (1.1,0) node[right] {$t$};
 \foreach \x/\l in {-1/{-1},0/{0},0.125/{\frac18},0.5/{\frac12},1/{1}}
  {\draw (\x,0.07) -- (\x,-0.07) node[below] {$\l$};}
 \draw (-1,0.45) -- (-1,0.55) -- (0.125,0.55) -- (0.125,0.45);
 \node[above] at (-0.4375,0.55) {PPT and separable};
 \draw (0.5,0.45) -- (0.5,0.55) -- (1,0.55) -- (1,0.45);
 \node[above,align=center] at (0.75,0.55) {distillable\\[-1pt]from one copy};
 \draw (0.125,1.1) -- (0.125,1.2); \draw (0.5,1.1) -- (0.5,1.2);
 \draw[densely dashed] (0.125,1.2) -- (0.5,1.2);
 \node[above,align=center] at (0.3125,1.2) {conjectured\\[-1pt]bound entangled};
 \draw (0.125,2.0) -- (0.125,2.1) -- (1,2.1) -- (1,2.0);
 \node[above] at (0.5625,2.1) {NPT};
 \draw[gray] (0.12,-0.02) -- (0.12,0.02);
 \draw[gray] (0.18,-0.02) -- (0.18,0.02);
 \draw[gray,densely dotted] (0.12,-0.55) -- (-0.9,-1.25);
 \draw[gray,densely dotted] (0.18,-0.55) -- (0.95,-1.25);
 \fill[green!22] (-0.9,-1.4) rectangle (-0.746,-1.25);
 \fill[orange!25] (-0.746,-1.4) rectangle (0.95,-1.25);
 \draw[thick] (-0.9,-1.4) -- (0.95,-1.4);
 \draw[line width=3pt] (-0.746,-1.13) -- (0.860,-1.13);
 \draw[fill=white] (-0.746,-1.13) circle (2.2pt);
 \fill (0.860,-1.13) circle (2.2pt);
 \node[above,fill=white,inner sep=1pt] at (0.057,-1.08) {bound entangled (Thm.~\ref{thm:family})};
 \fill (-0.664,-1.4) circle (1.8pt);
 \foreach \x/\l in {-0.746/{\frac18},-0.664/{\frac6{47}},0.860/{\frac{4427}{25000}}}
  {\draw (\x,-1.33) -- (\x,-1.47) node[below] {$\l$};}
 \node[below] at (-0.664,-1.95) {state of Thm.~\ref{thm:examples}(b)};
 \node[below,anchor=north east,inner xsep=0pt] at (0.95,-1.95) {state of Table~\ref{tab:intervals} for $d\ge8$};
\end{tikzpicture}
\caption{\label{fig:line}The line of Werner states $(I-t\,\swap)/(d^2-td)$ on $\mathbb C^d\otimes\mathbb C^d$, drawn to scale for $d=8$. The states with $t\le1/d$ are PPT and separable. The states with $t>1/d$ are NPT, and those with $t>1/2$ can be distilled from one copy. The states with $1/d<t\le1/2$ are conjectured to be bound entangled~\cite{divincenzo,dur}. The lower axis magnifies the interval $0.12\le t\le0.18$. Theorem~\ref{thm:family} proves the conjecture for the states just beyond the boundary. For $d=8$ it covers $1/8<t\le4427/25000$. The right end of this interval is the state $(25000\,I-4427\,\swap)/1564584$ of Table~\ref{tab:intervals}, whose proof uses a computer. The state $(47\,I-6\,\swap)/2960$ of Theorem~\ref{thm:examples}(b) has $t=6/47$, where the boundary has $t=6/48$.}
\end{figure}
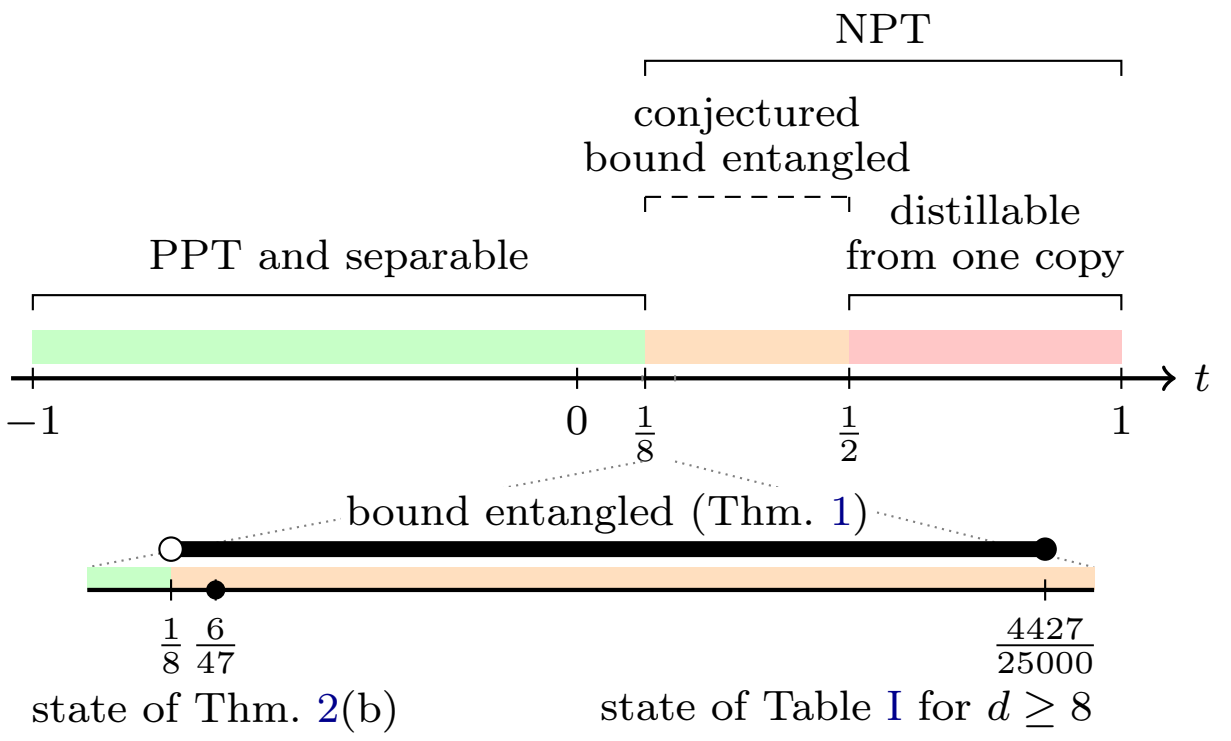

\step{Werner states and the PPT boundary} A Werner state on $\mathbb C^d\otimes\mathbb C^d$ is a state that commutes with $U\otimes U$ for every unitary $U$~\cite{werner}. Every such state is proportional to
\[
 I-t\,\swap\qquad\text{with }-1\le t\le1 .
\]
In Sec.~\ref{sec:higher} we show that its partial transpose has the eigenvalue $1-td$, up to a positive factor, on the vector $\sum_i\ket{i,i}$, and that all its other eigenvalues are positive. So the state is PPT exactly when $t\le1/d$. Werner showed that the PPT Werner states are separable~\cite{werner}. At the other end, the Werner states with $t>1/2$ can be distilled from a single copy~\cite{divincenzo,dur}. The states in between are NPT, but they cannot be distilled from one copy, and the conjecture of Refs.~\cite{divincenzo,dur} is that they are bound entangled. Figure~\ref{fig:line} shows this line of states.

\step{Building states at the boundary} Our states lie just beyond the boundary $t=1/d$. The states on the boundary cannot be distilled, for a simple reason: they are PPT. A state slightly beyond the boundary is close to such a state, but it is NPT, and so this reason no longer applies.

In our method, we write the state as the image of a maximally entangled vector under a product $L\otimes M$ of two local maps. The input spaces of the maps may be larger than their outputs. The proof is one matrix $Q$ for a single copy that satisfies four conditions: one linear equation and three matrix inequalities. These conditions imply that the state is undistillable for every number of copies. When the solution is strict, a nearby matrix satisfies the four conditions for all maps that are close enough to the original ones (Lemma~\ref{lem:stable}). So we look for local maps that contain a noise parameter, and whose output is close to the boundary. We lower the noise by a small amount, so that the output crosses the boundary, and we look for $Q$. For fixed maps, finding $Q$ is a semidefinite program, and a solution of this program can be checked in exact arithmetic. In dimension eight the step is small: the boundary state proportional to $48\,I-6\,\swap$ becomes the state proportional to $47\,I-6\,\swap$.

\step{Results} For a state $\omega$ on $\mathbb C^d\otimes\mathbb C^d$ we write
\[
 p(\omega)=-\tr(\swap\,\omega).
\]
So $-p(\omega)$ is the expectation value of the swap operator. In Sec.~\ref{sec:higher} we show that a Werner state $\omega$ is NPT exactly when $p(\omega)>0$.

\begin{theorem}
\label{thm:family}
Let $d\ge3$, and let $\omega$ be a Werner state on $\mathbb C^d\otimes\mathbb C^d$ with
\[
 0<p(\omega)\le p_0(d),
\]
where $p_0(d)$ is given in Table~\ref{tab:intervals}. Then $\omega$ is an NPT state, and its distillable entanglement under two-way LOCC is zero. Equivalently, by the distillability criterion of Ref.~\cite{horodecki98}, for every $n$, every two-qubit state that a separable operation produces from $\omega^{\otimes n}$ with nonzero probability is separable. In particular, each such two-qubit state has fidelity at most $1/2$ with a Bell state.
\end{theorem}

\begin{table}[t]
\caption{\label{tab:intervals}The values $p_0(d)$ in Theorem~\ref{thm:family}. Each value is $p(\omega)$ for an NPT bound entangled Werner state $\omega$ on $\mathbb C^{d_0}\otimes\mathbb C^{d_0}$. The last column gives $\omega$ up to normalization. The proofs are in Sec.~\ref{sec:computer}.}
\begin{ruledtabular}
\begin{tabular}{llcl}
$d$ & $p_0(d)$ & $d_0$ & State $\omega$\\
\hline
$3,4$ & $25/2097$ & $3$ & $1579\,I-543\,\swap$\\
$5,6$ & $1/55$ & $5$ & $23\,I-5\,\swap$\\
$7$ & $1/29$ & $7$ & $17\,I-3\,\swap$\\
$\ge8$ & $496/9313$ & $8$ & $25000\,I-4427\,\swap$\\
\end{tabular}
\end{ruledtabular}
\end{table}

The proofs of Theorem~\ref{thm:family} use a computer for the last step. We also give states whose proofs need less.

\begin{theorem}
\label{thm:examples}
The following Werner states are NPT, and they have the property of Theorem~\ref{thm:family}: separable operations produce only separable two-qubit states from their copies.
\begin{enumerate}[(a)]
 \item For every $d\ge128$, the state $\big((d-1)\,I-\swap\big)/\big((d-1)d^2-d\big)$ on $\mathbb C^d\otimes\mathbb C^d$.
 \item The state $(47\,I-6\,\swap)/2960$ on $\mathbb C^8\otimes\mathbb C^8$.
 \item The state $(1579\,I-543\,\swap)/12582$ on $\mathbb C^3\otimes\mathbb C^3$.
\end{enumerate}
\end{theorem}

The three proofs differ in how the last step is checked. For (a) the last step is a pair of norm estimates, and no computer is needed (Sec.~\ref{sec:large}). For (b) it is a computation with matrices of size at most three, which can be done by hand (Sec.~\ref{sec:eight}). The first version of this paper gave a longer proof of the same kind for the state $(23\,I-3\,\swap)/1448$. For (c) the last step is a computation in exact arithmetic on many small matrices, which needs a computer (Sec.~\ref{sec:qutrits}). Tabia and Hsieh gave a two-qutrit example~\cite{tabiahsieh}: they applied the method of the first version of this paper with larger input spaces. Brand\~ao has obtained, with the same method, two-qutrit Werner states much farther from the boundary than theirs, such as $(20000\,I-6877\,\swap)/159369$ and, by mixing, $(32\,I-11\,\swap)/255$, and he showed that their distillable secret key is zero as well~\cite{brandao}. We obtained the state (c) by extending his certificate.

We explain the terms in Theorem~\ref{thm:family}. A \emph{separable operation} is a completely positive map whose Kraus operators all have the product form $A\otimes B$, where $A$ acts on Alice's systems and $B$ acts on Bob's systems. We do not require that the map preserves the trace: every branch of an LOCC protocol, kept with postselection on its measurement outcomes, is such a map~\cite{locc}, and the trace of its output is then the probability of that branch. The output divided by its trace is the state in that case. The \emph{fidelity} of a two-qubit state $\tau$ with a Bell state $\ket\Phi$ is $\bra\Phi\tau\ket\Phi$. The value $1/2$ cannot be improved, because the product state $\ket{0,0}$ has fidelity $1/2$ with the Bell state $(\ket{0,0}+\ket{1,1})/\sqrt2$.

In the notation of Ref.~\cite{fiveopen}, the state of Theorem~\ref{thm:examples}(b) is the Werner state $\rho(8,-\tfrac{6}{47})$. Theorem~\ref{thm:family} also gives the form of undistillability that is used there: for every $n$ and for all projectors $P_A$ on Alice's systems and $P_B$ on Bob's systems that have rank two, the operator $(P_A\otimes P_B)\,\omega^{\otimes n}(P_A\otimes P_B)$ has a positive partial transpose. Indeed, if we identify the ranges of $P_A$ and $P_B$ with qubits, this operator is either zero or a positive multiple of a two-qubit state that a separable operation produces from $\omega^{\otimes n}$. By Theorem~\ref{thm:family} this state is separable, so its partial transpose is positive semidefinite.

Our results also apply to states that are not Werner states. Let $\ket\varphi$ be a unit vector on $\mathbb C^{d_A}\otimes\mathbb C^{d_B}$, and let $D=d_Ad_B$. Bandyopadhyay and Roychowdhury consider the states~\cite{br03}
\begin{equation}\label{eq:br}
 \rho_\epsilon=\epsilon\,\sigma+(1-\epsilon)\,\frac{(I-\ket\varphi\!\bra\varphi)^{\Gamma}}{D-1},\qquad 0\le\epsilon\le1,
\end{equation}
where $\sigma$ is a state with $\bra\varphi\sigma^\Gamma\ket\varphi<0$. These states are NPT for $\epsilon>0$. Bandyopadhyay and Roychowdhury proved that for every fixed $n$ there is an $\epsilon_n>0$ such that $\rho_\epsilon$ cannot be distilled from $n$ copies when $0<\epsilon\le\epsilon_n$~\cite[Theorem~1]{br03}. In their concluding remarks they conjectured that these states are bound entangled, and that for each choice of $\ket\varphi$ and $\sigma$ there is one interval $0<\epsilon\le\epsilon_\infty$ on which $\rho_\epsilon$ is NPT and bound entangled~\cite{br03}. Their families assume that $\ket\varphi$ has Schmidt rank at least three. Their Theorem~2 extends the family to states built from several vectors $\ket{\varphi_i}$ in orthogonal subspaces~\cite[Theorem~2]{br03}. We prove their conjecture about the interval for the family~\eqref{eq:br} with one vector, for every choice of $\ket\varphi$ and $\sigma$. We do not treat the extension with several vectors.

\begin{theorem}
\label{thm:br}
Let $\ket\varphi$ be a unit vector of Schmidt rank at least three on $\mathbb C^{d_A}\otimes\mathbb C^{d_B}$, and let $\sigma$ be a state with $\bra\varphi\sigma^\Gamma\ket\varphi<0$. Then there is an $\epsilon_\infty>0$ such that the state $\rho_\epsilon$ of Eq.~\eqref{eq:br} is NPT and bound entangled for every $0<\epsilon\le\epsilon_\infty$.
\end{theorem}

The Schmidt rank must be at least three: if $\ket\varphi$ has Schmidt rank two, then $\bra\varphi\rho_\epsilon^\Gamma\ket\varphi<0$ implies that $\rho_\epsilon$ is distillable from one copy~\cite{horodecki98,dur}.

\step{Consequences} The existence of NPT bound entangled states has a known consequence~\cite[Sec.~III~A]{fiveopen}. A function $E$ of states is \emph{additive} if $E(\rho_1\otimes\rho_2)=E(\rho_1)+E(\rho_2)$ for all states $\rho_1$ and $\rho_2$. It is \emph{convex} if
\[
 E\big(q\rho_1+(1-q)\rho_2\big)\le q\,E(\rho_1)+(1-q)\,E(\rho_2)
\]
for all states $\rho_1$ and $\rho_2$ and all $0\le q\le1$. In words, the value of $E$ on a mixture of two states is at most the same mixture of its two values. Let $\rho$ be an NPT bound entangled state. Every NPT state is distillable by PPT-preserving operations~\cite{eggeling}, and it becomes distillable by LOCC when a suitable PPT state $\sigma$ is added~\cite{vollbrecht}. So $E_D(\rho)=E_D(\sigma)=0$ while $E_D(\rho\otimes\sigma)>0$, and additivity fails. This strong failure of additivity, in which two states that are useless alone are useful together, is called \emph{superactivation}.

Convexity fails as well, by an argument of Shor, Smolin and Terhal~\cite[Eqs.~(3)--(5)]{sst}. We give Alice one extra qubit as a flag, and we consider the two states $\rho'=\rho\otimes\ket0\!\bra0$ and $\sigma'=\sigma\otimes\ket1\!\bra1$. If $\rho$ and $\sigma$ act on different spaces, we first embed both in a common space. The flag is a local system in a fixed state, so $E_D(\rho')=E_D(\rho)=0$ and $E_D(\sigma')=E_D(\sigma)=0$. Now suppose that Alice and Bob share many copies of the equal mixture $\tfrac12\rho'+\tfrac12\sigma'$. Alice measures the flags and tells Bob the outcomes. Then both know which copies are copies of $\rho$ and which are copies of $\sigma$, and about half of the copies are of each kind. They form pairs $\rho\otimes\sigma$ and distill them. So the mixture has positive distillable entanglement, although $E_D(\rho')$ and $E_D(\sigma')$ are zero. This contradicts the inequality above.

\begin{corollary}
Distillable entanglement is neither additive nor convex.
\end{corollary}

\step{Structure of the paper} Section~\ref{sec:prelims} fixes the notation and recalls standard facts about completely positive maps and Choi matrices. In Sec.~\ref{sec:higher} we show how one NPT bound entangled state gives NPT bound entangled Werner states in its own dimension and in every larger one. So one such state in dimension three already gives NPT bound entangled states in every dimension $d\ge3$. Section~\ref{sec:method} contains the method. Its lemmas do not depend on the dimension. Together they show that a solution of a semidefinite program for one copy certifies undistillability for every number of copies. The same argument shows that the distillable secret key is zero (Proposition~\ref{prop:key}). We also explain how a solution of this program is checked, and we describe the accompanying program that checks our certificates. The next three sections prove Theorem~\ref{thm:examples}: in Sec.~\ref{sec:large} without a computer for $d\ge128$, in Sec.~\ref{sec:eight} by hand in dimension eight, and in Sec.~\ref{sec:computer} with a computer, where we also prove Theorem~\ref{thm:family}. In Sec.~\ref{sec:nonwerner} we go beyond Werner states. There we prove Theorem~\ref{thm:br}, we give a ball in which all states but one are NPT and bound entangled, and we give the two-qutrit state
\[
 \frac{31\,I-14\,\swap+14\,\ket{2,2}\!\bra{2,2}+28\,\ket{3,3}\!\bra{3,3}}{279} ,
\]
which we obtain from a two-qutrit Werner state by local filters and separable noise. The longer computations are in the appendices.

\section{Notation and standard facts}
\label{sec:prelims}

\step{Notation} We write $X^{\dagger}$ for the conjugate transpose of a matrix $X$, and $X\succeq0$ if $X$ is positive semidefinite. We write vectors as kets. The bra $\bra{v}$ is the conjugate transpose of the ket $\ket{v}$, so $\ket{v}\!\bra{w}$ is a matrix and $\langle v|w\rangle$ is a number. We write $\ket1,\dots,\ket d$ for the standard basis of $\mathbb C^d$. In sums over this basis, the indices run from $1$ to $d$. For qubits we use the labels $0$ and $1$. We let $\ket{u_d}=\sum_{i=1}^{d}\ket{i,i}$ in $\mathbb C^d\otimes\mathbb C^d$. This is a maximally entangled vector, not normalized. Let $T$ be the transpose, $T(X)=X^{T}$.

\step{Completely positive maps} A linear map $\Lambda$ on matrices is \emph{completely positive} if it preserves positive semidefinite block matrices: whenever a block matrix with the blocks $G_{\alpha\beta}$ is positive semidefinite, so is the block matrix with the blocks $\Lambda(G_{\alpha\beta})$. The map $\Lambda$ is \emph{completely copositive} if $\Lambda\circ T$ is completely positive.

\step{Choi matrices} The \emph{Choi matrix} of a linear map $\Lambda$ from matrices on $\mathbb C^{k}$ to matrices on $\mathbb C^{m}$ is
\[
 J_\Lambda=\sum_{p,q}\ket p\!\bra q\otimes\Lambda(\ket p\!\bra q),
\]
where $\ket p$ and $\ket q$ run over the standard basis of $\mathbb C^k$. It is a matrix on $\mathbb C^k\otimes\mathbb C^m$. We call $\mathbb C^k$ the \emph{input} and $\mathbb C^m$ the \emph{output}. For a matrix $X$ on $\mathbb C^k\otimes\mathbb C^m$ and basis vectors $\ket p,\ket q$ of the input, the \emph{block} $X_{pq}=(\bra p\otimes I)X(\ket q\otimes I)$ is a matrix on the output. So $(J_\Lambda)_{pq}=\Lambda(\ket p\!\bra q)$. We write $X^{\Gin}$ for the transpose of $X$ on the input: $(X^{\Gin})_{pq}=X_{qp}$. We use four standard facts.

\begin{fact}[Choi matrices]\label{fact:choi}
The following statements hold for linear maps from matrices on $\mathbb C^k$ to matrices on $\mathbb C^m$.
\begin{enumerate}[(a)]
 \item\label{fact:choi-a} Every matrix on $\mathbb C^k\otimes\mathbb C^m$ is the Choi matrix of exactly one such map~\cite[Sec.~2.2.2, Eq.~(2.65)]{watrous}.
 \item\label{fact:choi-b} A linear map is completely positive exactly when its Choi matrix is positive semidefinite~\cite[Theorem~2.22]{watrous}.
 \item\label{fact:choi-c} A linear map $\Lambda$ satisfies $\Lambda(X^{\dagger})=\Lambda(X)^{\dagger}$ for all $X$ exactly when its Choi matrix is Hermitian~\cite[Theorem~2.25]{watrous}.
 \item\label{fact:choi-d} The Choi matrix of a tensor product of maps is the tensor product of their Choi matrices, after the systems are put in order. As a formula, for maps $\Lambda_i$ from matrices on $\mathbb C^{k_i}$ to matrices on $\mathbb C^{m_i}$, with $i=1,\dots,n$,
 \[
  J_{\Lambda_1\otimes\dots\otimes\Lambda_n}=W\,\big(J_{\Lambda_1}\otimes\dots\otimes J_{\Lambda_n}\big)\,W^{\dagger} ,
 \]
 where $W$ is the permutation of the $2n$ tensor factors that moves the $n$ input systems in front of the $n$ output systems. In particular, $J_{\Lambda^{\otimes n}}=W\,J_\Lambda^{\otimes n}\,W^{\dagger}$.
\end{enumerate}
\end{fact}

Fact~\ref{fact:choi}(\ref{fact:choi-d}) follows directly from the definition. In the definition of Ref.~\cite[Eq.~(2.65)]{watrous}, the output systems come before the input systems. This changes only the order of the systems.

By the definition of the Choi matrix, the Choi matrix of $\Lambda\circ T$ is $J_\Lambda^{\Gin}$. So by Fact~\ref{fact:choi}(\ref{fact:choi-b}), a map $\Lambda$ is completely copositive exactly when $J_\Lambda^{\Gin}\succeq0$.

\section{From one dimension to all larger dimensions}
\label{sec:higher}

\step{The ratio that we have to pass} Consider a state proportional to $a\,I-b\,\swap$ on $\mathbb C^d\otimes\mathbb C^d$, with $a>0$. We have
\begin{align*}
 \swap&=\sum_{i,j=1}^{d}\ket j\!\bra i\otimes\ket i\!\bra j ,\\
 \swap^\Gamma&=\sum_{i,j=1}^{d}\ket j\!\bra i\otimes\ket j\!\bra i=\ket{u_d}\!\bra{u_d} ,
\end{align*}
because the partial transpose transposes the second factor. So
\[
 (a\,I-b\,\swap)^\Gamma=a\,I-b\,\ket{u_d}\!\bra{u_d}.
\]
Since $\langle u_d|u_d\rangle=d$, this matrix has the eigenvalue $a-bd$ on $\ket{u_d}$. Its other eigenvalue is $a$. So the state is NPT exactly when
\begin{equation}\label{eq:ratio}
 \frac ba>\frac1d .
\end{equation}
So the ratio $b/a$ of the two coefficients has to be larger than $1/d$. In the notation of the Introduction, $t=b/a$.

\step{The parameter $p$} For a state $\tau$ on $\mathbb C^d\otimes\mathbb C^d$ we have
\begin{equation}\label{eq:witness}
 \bra{u_d}\tau^\Gamma\ket{u_d}=\tr\big(\swap\,\tau\big)=-p(\tau),
\end{equation}
because $\swap^\Gamma=\ket{u_d}\!\bra{u_d}$. So every state with $p(\tau)>0$ is NPT. For the Werner state proportional to $I-t\,\swap$ we have $\tr\big(\swap\,(I-t\,\swap)\big)=d\,(1-td)$, and its trace is $d\,(d-t)$. So
\begin{equation}\label{eq:pt}
 p=\frac{td-1}{d-t},\qquad\text{and}\qquad t=\frac{1+dp}{d+p}.
\end{equation}
In particular, a Werner state is NPT exactly when $p>0$, and it is determined by its value of $p$. For the state of Theorem~\ref{thm:examples}(b) we have $p=1/370$.

\step{Three local operations} An LOCC operation cannot turn an undistillable state into a distillable one, because a distillation protocol for the result would give a distillation protocol for the original state. The same holds for an LOCC operation that succeeds only with some positive probability, provided that Alice and Bob know whether it succeeded.

\begin{lemma}[From one state to a family]\label{lem:ops}
Let $\tau$ be an undistillable state on $\mathbb C^{d'}\otimes\mathbb C^{d'}$ with $p(\tau)>0$, and let $d\ge d'$. Then for every $p$ with $0<p\le p(\tau)$, the Werner state on $\mathbb C^d\otimes\mathbb C^d$ with the parameter $p$ is NPT and bound entangled.
\end{lemma}

\begin{proof}
Alice and Bob apply the following three operations by LOCC with shared randomness.

\emph{Step 1: embedding.} Both apply the same isometry $V$ from $\mathbb C^{d'}$ into $\mathbb C^d$. Since $\swap\,(V\otimes V)=(V\otimes V)\,\swap$, the value of $p$ does not change.

\emph{Step 2: twirling.} Both apply the same random unitary $U$, chosen from the Haar measure. The result is the average of $(U\otimes U)\,\tau\,(U\otimes U)^\dagger$ over all unitaries, which is a Werner state~\cite{werner}. Since $\swap$ commutes with $U\otimes U$, the value of $p$ does not change.

\emph{Step 3: mixing.} Let $\pi_d$ be the result of twirling the product state $\ket{1,2}\!\bra{1,2}$. It is a separable Werner state with $p(\pi_d)=0$. With probability $1-\lambda$, Alice and Bob replace their state by $\pi_d$. This multiplies $p$ by $\lambda$.

A Werner state on $\mathbb C^d\otimes\mathbb C^d$ is determined by its value of $p$. So the three steps produce every Werner state with $0<p\le p(\tau)$, and each of these states is undistillable. Each is NPT by \eqref{eq:witness}, hence entangled~\cite{peres}.
\end{proof}

\step{Every NPT state} The lemma assumes that $p(\tau)>0$. Not every NPT state has this property, but a local filter turns every NPT state $\tau$ on $\mathbb C^{d'}\otimes\mathbb C^{d'}$ into a state that has it. There is a vector $\ket\psi$ with $\bra\psi\tau^\Gamma\ket\psi<0$, and we write it as $\ket\psi=(A\otimes I)\ket{u_{d'}}$ for a matrix $A$ on $\mathbb C^{d'}$. The partial transpose acts on Bob's system, so $\big((A^\dagger\otimes I)\,\tau\,(A\otimes I)\big)^\Gamma=(A^\dagger\otimes I)\,\tau^\Gamma(A\otimes I)$. So by \eqref{eq:witness}, the normalized state $\tau'$ proportional to $(A^\dagger\otimes I)\,\tau\,(A\otimes I)$ has $p(\tau')>0$. Alice obtains $\tau'$ from $\tau$ with the filter $A^\dagger$, rescaled to be a Kraus operator of a local measurement. So $\tau'$ is undistillable if $\tau$ is, and the lemma applies to it.

\step{The threshold} The undistillable NPT Werner states form one interval in each dimension. A state $\tau$ is distillable exactly when, for some $n$, there is a vector $\ket\psi$ of Schmidt rank two with $\bra\psi(\tau^{\otimes n})^\Gamma\ket\psi<0$~\cite{horodecki98,divincenzo,dur}. For $n=1$ and a Werner state with $t\ge0$, which includes all NPT Werner states, the smallest value of $\bra\psi(I-t\ket{u_d}\!\bra{u_d})\ket\psi$ over unit vectors of Schmidt rank two is $1-2t$. So a Werner state is distillable from one copy exactly when $t>1/2$.

\begin{theorem}[The threshold]\label{thm:threshold}
For every $d\ge2$ there is a number $p_*(d)\ge0$ such that a Werner state on $\mathbb C^d\otimes\mathbb C^d$ is NPT and bound entangled exactly when $0<p\le p_*(d)$. Let $t_*(d)$ be the value of $t$ that corresponds to $p_*(d)$ in \eqref{eq:pt}. Then:
\begin{enumerate}[(i)]
 \item $p_*(d)\ge p_0(d)$ for $d\ge3$, and $p_*(d)\le(d-2)/(2d-1)$ for $d\ge2$.
 \item $p_*(d)$ does not decrease when $d$ grows.
 \item $t_*(d)$ does not increase when $d$ grows.
 \item Both sequences have the same limit $c$, and $c\le t_*(d)\le(1+dc)/(d+c)$.
\end{enumerate}
The conjecture of Refs.~\cite{divincenzo,dur} is equivalent to $c=1/2$.
\end{theorem}

\begin{proof}
Let $P_d$ be the set of all $p>0$ for which the Werner state is undistillable. By the mixing step in the proof of Lemma~\ref{lem:ops}, $P_d$ contains with every $p$ all smaller positive numbers. The criterion above describes the undistillable states by the inequalities $\bra\psi(\tau^{\otimes n})^\Gamma\ket\psi\ge0$, so the set of undistillable states is closed, and $P_d$ contains its supremum $p_*(d)$. If $P_d$ is empty we let $p_*(d)=0$.

(i) The lower bound is Theorem~\ref{thm:family}. A Werner state with $t>1/2$ is distillable, and $t=1/2$ corresponds to $p=(d-2)/(2d-1)$ in \eqref{eq:pt}.

(ii) Embedding and twirling keep $p$, and they lead from dimension $d$ to dimension $d+1$.

(iii) Let Alice and Bob both project onto the same subspace of dimension $d$ in $\mathbb C^{d+1}$. This is a local operation that succeeds with positive probability, and it turns $I-t\,\swap$ in dimension $d+1$ into $I-t\,\swap$ in dimension $d$. So if the first state is undistillable, the second one is too. We apply this with $t=t_*(d+1)$. If the resulting state in dimension $d$ is NPT, then $t\le t_*(d)$. If it is PPT, then $t\le1/d\le t_*(d)$.

(iv) By \eqref{eq:pt}, $t_*(d)-p_*(d)=(1-p_*(d)^2)/(d+p_*(d))$, which tends to zero. So the two monotone sequences have the same limit $c$, with $p_*(d)\le c\le t_*(d)$. The upper bound on $t_*(d)$ follows from $p_*(d)\le c$ and \eqref{eq:pt}.

The conjecture says that $t_*(d)=1/2$ for every $d\ge3$. Since $c\le t_*(d)\le1/2$, this holds exactly when $c=1/2$.
\end{proof}

At present we know that $496/9313\le c\le1/2$. The lower bound is shown at the end of Sec.~\ref{sec:computer}.

\section{The method}
\label{sec:method}

The lemmas of this section are used in all our proofs. We state them for the following data.
\begin{itemize}
 \item An input space $\mathcal H=\mathbb C^r$ with the standard basis, and the vector $\ket u=\sum_{a=1}^{r}\ket{a,a}$ in $\mathcal H\otimes\mathcal H$. Alice holds the first factor and Bob holds the second.
 \item Two completely positive maps $L$ and $M$ from matrices on $\mathcal H$ to matrices on $\mathbb C^d$, and their product $\mathcal C=L\otimes M$.
 \item The matrix $R=\mathcal C(\ket u\!\bra u)$ on $\mathbb C^d\otimes\mathbb C^d$, which we assume to be invertible, and the state $\rho=R/\tr(R)$.
\end{itemize}
The input space need not be $\mathbb C^d$. In Secs.~\ref{sec:large} and~\ref{sec:eight} it is $\mathbb C^d$, but for the states in low dimensions we found $Q$ only with larger input spaces, as Tabia and Hsieh also observed~\cite{tabiahsieh}. For $n$ copies we write $\mathcal C_n=\mathcal C^{\otimes n}$, $\ket{u_n}=\ket u^{\otimes n}$ and $R_n=R^{\otimes n}$, so that $R_n=\mathcal C_n(\ket{u_n}\!\bra{u_n})$ and $\rho^{\otimes n}=R_n/\tr(R)^n$.

We write $J$ for the Choi matrix of $\mathcal C$ (Sec.~\ref{sec:prelims}). It is a matrix on four systems, which we order as Alice's input, Bob's input, Alice's output and Bob's output. Its input is $\mathcal H\otimes\mathcal H$, its output is $\mathbb C^d\otimes\mathbb C^d$, and its blocks are $J_{pq}=\mathcal C(\ket p\!\bra q)$. For a matrix $X$ on these four systems, $X^{\Gin}$ is the transpose on the whole input $\mathcal H\otimes\mathcal H$.

\subsection{Four conditions for one copy}

Let $Q$ be a Hermitian matrix on the same four systems as $J$. By Fact~\ref{fact:choi}(\ref{fact:choi-a}), $Q$ is the Choi matrix of exactly one linear map $\mathcal K$ from matrices on $\mathcal H\otimes\mathcal H$ to matrices on $\mathbb C^d\otimes\mathbb C^d$. Let $V=\ket u\otimes I$, where $I$ is the identity on the output, so that $V\ket x=\ket u\otimes\ket x$. We consider the conditions
\begin{subequations}\label{eq:cert}
\begin{align}
 J-Q&\succeq0, \label{eq:cert-a}\\
 J+Q&\succeq0, \label{eq:cert-b}\\
 Q^{\Gin}&\succeq0, \label{eq:cert-c}\\
 (J-Q)\,V&=0 . \label{eq:cert-d}
\end{align}
\end{subequations}
If $\mathcal C$ were completely copositive, then $Q=J$ would satisfy all four conditions. For our maps this is not the case (see Sec.~\ref{sec:eight}), and we use another matrix $Q$ in the place of $J$. By \eqref{eq:cert-c}, $\mathcal K$ is completely copositive. By \eqref{eq:cert-d}, it agrees with $\mathcal C$ on all matrices $\ket z\!\bra u$. By \eqref{eq:cert-a} and \eqref{eq:cert-b}, the maps $\mathcal C-\mathcal K$ and $\mathcal C+\mathcal K$ are completely positive.

\begin{lemma}[One copy implies every number of copies]\label{lem:reverse}
Suppose that $Q$ satisfies \eqref{eq:cert}. Then for every $n\ge1$ and every vector $\ket z$ in $(\mathcal H\otimes\mathcal H)^{\otimes n}$, the space that contains $\ket{u_n}$, the matrix
\begin{equation}\label{eq:reverse}
 \Omega_n(z)=\begin{pmatrix}
  R_n & \mathcal C_n(\ket z\!\bra{u_n})\\[2pt]
  \mathcal C_n(\ket{u_n}\!\bra z) & \mathcal C_n(\ket z\!\bra z)
 \end{pmatrix}
\end{equation}
is positive semidefinite.
\end{lemma}

Complete positivity of $\mathcal C_n$ gives the same block matrix with the two blocks off the diagonal exchanged. The content of the lemma is that the matrix stays positive semidefinite in the order \eqref{eq:reverse}. We call $\Omega_n(z)$ the \emph{reversed block matrix}.

\begin{proof}
Let $\mathcal K_n=\mathcal K^{\otimes n}$. Three facts follow from the conditions \eqref{eq:cert}, for every $n$.
\begin{enumerate}[(i)]
 \item\label{it:copositive} \emph{$\mathcal K_n$ is completely copositive.} The Choi matrix of $\mathcal K\circ T$ is $Q^{\Gin}$, which is positive semidefinite by \eqref{eq:cert-c}. So $\mathcal K\circ T$ is completely positive by Fact~\ref{fact:choi}(\ref{fact:choi-b}). Its tensor power is $\mathcal K_n\circ T$, and by Fact~\ref{fact:choi}(\ref{fact:choi-d}) the Choi matrix of $\mathcal K_n\circ T$ is a tensor power of $Q^{\Gin}$, after the systems are put in order. It is positive semidefinite, so $\mathcal K_n\circ T$ is completely positive as well.
 \item\label{it:agree} \emph{$\mathcal K_n$ and $\mathcal C_n$ agree on $\ket z\!\bra{u_n}$ and on $\ket{u_n}\!\bra z$, for every vector $\ket z$.} We start with one copy. Let $\ket p$ and $\ket q$ run over the standard basis of the input $\mathcal H\otimes\mathcal H$. By the definition of the Choi matrix,
 \[
  J-Q=\sum_{p,q}\ket p\!\bra q\otimes(\mathcal C-\mathcal K)(\ket p\!\bra q).
 \]
 We multiply this equation by $V=\ket u\otimes I$ from the right. In each term, $\bra q$ is applied to $\ket u$, and the result is the number $\langle q|u\rangle$. Since the numbers $\langle q|u\rangle$ are real, $\sum_q\langle q|u\rangle\,\ket p\!\bra q=\ket p\!\bra u$, and we get
 \[
  (J-Q)V=\sum_p\ket p\otimes(\mathcal C-\mathcal K)(\ket p\!\bra u).
 \]
 The left-hand side is zero by \eqref{eq:cert-d}. The vectors $\ket p$ are orthonormal, so each term on the right-hand side is zero. So $\mathcal K(\ket p\!\bra u)=\mathcal C(\ket p\!\bra u)$ for every basis vector $\ket p$, and by linearity $\mathcal K(\ket z\!\bra u)=\mathcal C(\ket z\!\bra u)$ for every vector $\ket z$.

 Next we exchange the two sides. The matrices $J$ and $Q$ are Hermitian, so by Fact~\ref{fact:choi}(\ref{fact:choi-c}) both maps satisfy $\Lambda(X^{\dagger})=\Lambda(X)^{\dagger}$. With $X=\ket z\!\bra u$ we get $\mathcal K(\ket u\!\bra z)=\mathcal C(\ket u\!\bra z)$.

 Finally we pass to $n$ copies. For basis vectors $\ket{p_1},\dots,\ket{p_n}$ of the input,
 \[
  \mathcal K_n\big(\ket{p_1,\dots,p_n}\!\bra{u_n}\big)=\bigotimes_{k=1}^{n}\mathcal K(\ket{p_k}\!\bra u)=\bigotimes_{k=1}^{n}\mathcal C(\ket{p_k}\!\bra u),
 \]
 and the right-hand side is $\mathcal C_n(\ket{p_1,\dots,p_n}\!\bra{u_n})$. Every $\ket z$ is a linear combination of the vectors $\ket{p_1,\dots,p_n}$, so the two maps agree on $\ket z\!\bra{u_n}$. As for one copy, they also agree on $\ket{u_n}\!\bra z$.
 \item\label{it:difference} \emph{$\mathcal C_n-\mathcal K_n$ is completely positive.} By Fact~\ref{fact:choi}(\ref{fact:choi-d}), its Choi matrix is $J^{\otimes n}-Q^{\otimes n}$, after the systems are put in order. Let $J_+=(J+Q)/2$ and $J_-=(J-Q)/2$. These matrices are positive semidefinite by \eqref{eq:cert-a} and \eqref{eq:cert-b}, and $J=J_++J_-$, $Q=J_+-J_-$. We expand both tensor powers:
 \begin{align*}
  J^{\otimes n}&=\sum_{c}J_{c_1}\otimes\dots\otimes J_{c_n},\\
  Q^{\otimes n}&=\sum_{c}(-1)^{m(c)}\,J_{c_1}\otimes\dots\otimes J_{c_n}.
 \end{align*}
 Here $c=(c_1,\dots,c_n)$ runs over the $2^n$ strings of signs $+$ and $-$, and $m(c)$ is the number of minus signs in $c$. In the difference, the terms with even $m(c)$ cancel, and the terms with odd $m(c)$ appear twice:
 \[
  J^{\otimes n}-Q^{\otimes n}=2\sum_{c:\,m(c)\text{ odd}}J_{c_1}\otimes\dots\otimes J_{c_n}.
 \]
 Each term is a tensor product of positive semidefinite matrices, so it is positive semidefinite. So $\mathcal C_n-\mathcal K_n$ is completely positive by Fact~\ref{fact:choi}(\ref{fact:choi-b}).
\end{enumerate}
Now fix $\ket z$. The block matrix
\[
 \begin{pmatrix}
  \ket{u_n}\!\bra{u_n}&\ket{u_n}\!\bra{z}\\
  \ket{z}\!\bra{u_n}&\ket{z}\!\bra{z}
 \end{pmatrix}
\]
is positive semidefinite, and so is its transpose,
\[
 \begin{pmatrix}
  (\ket{u_n}\!\bra{u_n})^{T}&(\ket{z}\!\bra{u_n})^{T}\\
  (\ket{u_n}\!\bra{z})^{T}&(\ket{z}\!\bra{z})^{T}
 \end{pmatrix}.
\]
By (\ref{it:copositive}), the map $\mathcal K_n\circ T$ is completely positive. We apply it to the four blocks of the transpose. The result
\[
 \begin{pmatrix}
  \mathcal K_n(\ket{u_n}\!\bra{u_n})&\mathcal K_n(\ket{z}\!\bra{u_n})\\
  \mathcal K_n(\ket{u_n}\!\bra{z})&\mathcal K_n(\ket{z}\!\bra{z})
 \end{pmatrix}
\]
is positive semidefinite. By (\ref{it:agree}), its first three blocks do not change when we replace $\mathcal K_n$ by $\mathcal C_n$. By (\ref{it:difference}), the matrix stays positive semidefinite when we add $(\mathcal C_n-\mathcal K_n)(\ket{z}\!\bra{z})$ to the last block. The result is $\Omega_n(z)$, since $R_n=\mathcal C_n(\ket{u_n}\!\bra{u_n})$.
\end{proof}

\subsection{From the reversed block inequality to undistillability}

\begin{lemma}[Undistillability]\label{lem:fidelity}
Suppose that the reversed block matrix $\Omega_n(z)$ of \eqref{eq:reverse} is positive semidefinite for every $n$ and every $\ket z$. Then the distillable entanglement of $\rho$ is zero. Moreover, for every $n$, every two-qubit state that a separable operation produces from $\rho^{\otimes n}$ with nonzero probability is separable, and its fidelity with a Bell state is at most $1/2$.
\end{lemma}

The proof adapts the proof of Lemma~3.2 of Ref.~\cite{openai}, which assumes that the two local maps are completely copositive. We use the inequality \eqref{eq:reverse} in place of this assumption.

\begin{proof}
\emph{Step 1: one Kraus operator.} Let $\ket0,\ket1$ be the standard basis of a qubit, and let $\ket\Phi=(\ket{0,0}+\ket{1,1})/\sqrt2$. For $n$ copies we group Alice's $n$ systems together, and Bob's as well, so that $\mathcal C_n=L^{\otimes n}\otimes M^{\otimes n}$. A separable operation is a sum of terms $X\mapsto(K_A\otimes K_B)X(K_A\otimes K_B)^\dagger$. Consider one term whose output is a pair of qubits. We absorb $K_A$ and $K_B$ into the maps: let
\begin{align*}
 \widetilde L(X)&=K_A\,L^{\otimes n}(X)\,K_A^\dagger,\\
 \widetilde M(X)&=K_B\,M^{\otimes n}(X)\,K_B^\dagger,
\end{align*}
and $\widetilde{\mathcal C}=\widetilde L\otimes\widetilde M$. Since $\rho^{\otimes n}$ is a positive multiple of $R_n$, the output of this term on $\rho^{\otimes n}$ is a positive multiple of $\tau=\widetilde{\mathcal C}(\ket{u_n}\!\bra{u_n})$. The map $\widetilde{\mathcal C}$ is completely positive. For every $\ket z$, the reversed block matrix of $\widetilde{\mathcal C}$, with the blocks $\tau$, $\widetilde{\mathcal C}(\ket z\!\bra{u_n})$, $\widetilde{\mathcal C}(\ket{u_n}\!\bra z)$ and $\widetilde{\mathcal C}(\ket z\!\bra z)$, is positive semidefinite, because it is the congruence of $\Omega_n(z)$ with $K_A\otimes K_B$ in both blocks. We write
\[
 p_{ij}=\bra{i,j}\tau\ket{i,j},\qquad c=\bra{0,0}\tau\ket{1,1} .
\]

\emph{Step 2: the operators $E_i$ and $F_j$.} Let $\widetilde L^\dagger$ be the adjoint of $\widetilde L$ for the trace inner product, so that $\tr(\widetilde L(X)\,Y)=\tr(X\,\widetilde L^\dagger(Y))$, and define $\widetilde M^\dagger$ in the same way. Let
\[
 E_i=\widetilde L^\dagger(\ket i\!\bra i),\qquad F_j=\widetilde M^\dagger(\ket j\!\bra j).
\]
These matrices are positive semidefinite. For every matrix $Y$ on Alice's input and $\ket z=(Y\otimes I)\ket{u_n}$,
\begin{equation}\label{eq:pull}
 \bra{i,j}\widetilde{\mathcal C}(\ket z\!\bra z)\ket{i,j}=\bra{u_n}\big(Y^\dagger E_iY\otimes F_j\big)\ket{u_n}.
\end{equation}
In particular, $p_{ij}=\bra{u_n}(E_i\otimes F_j)\ket{u_n}$.

Let $H=E_0+E_1$, let $H^{-1}$ be the inverse of $H$ on its support, extended by zero, and let $Y_0=H^{-1}E_0$. We claim that
\begin{equation}\label{eq:quarter}
 Y_0^\dagger E_1Y_0\preceq\tfrac14E_0,\qquad (I-Y_0)^\dagger E_0(I-Y_0)\preceq\tfrac14E_1 .
\end{equation}
To see this, let $X=H^{-1/2}E_0H^{-1/2}$. Then $0\preceq X\preceq I$, $E_0=H^{1/2}XH^{1/2}$, and $H^{-1/2}E_1H^{-1/2}$ equals $I-X$ on the support of $H$. So
\[
 Y_0^\dagger E_1Y_0=H^{1/2}X(I-X)XH^{1/2}\preceq\tfrac14H^{1/2}XH^{1/2},
\]
because $x^2(1-x)\le x/4$ for $0\le x\le1$. This is the first inequality. On the support of $H$ we have $I-Y_0=H^{-1}E_1$, and $E_0$ vanishes outside this support. So the second inequality is the first one with $E_0$ and $E_1$ exchanged.

Now let $\ket v=(Y_0\otimes I)\ket{u_n}$ and $\ket w=\ket{u_n}-\ket v$. By \eqref{eq:pull} and \eqref{eq:quarter},
\begin{equation}\label{eq:branches}
\begin{aligned}
 \bra{1,1}\widetilde{\mathcal C}(\ket v\!\bra v)\ket{1,1}&\le\tfrac14\,p_{01},\\
 \bra{0,0}\widetilde{\mathcal C}(\ket w\!\bra w)\ket{0,0}&\le\tfrac14\,p_{10}.
\end{aligned}
\end{equation}

\emph{Step 3: a fidelity bound smaller than one.} Since $\ket{u_n}=\ket v+\ket w$, the number $c$ is the sum of $\bra{0,0}\widetilde{\mathcal C}(\ket v\!\bra{u_n})\ket{1,1}$ and $\bra{0,0}\widetilde{\mathcal C}(\ket w\!\bra{u_n})\ket{1,1}$. We bound each of them by a $2\times2$ determinant. For the first one we use the reversed block matrix of Step~1 with $\ket z=\ket v$, evaluated on $\ket{0,0}$ in the first block and on $\ket{1,1}$ in the second block. Its diagonal entries are $p_{00}$ and $\bra{1,1}\widetilde{\mathcal C}(\ket v\!\bra v)\ket{1,1}$. For the second one we use the block matrix of $\widetilde{\mathcal C}$ on $\ket w$ and $\ket{u_n}$ in the usual order, which is positive semidefinite because $\widetilde{\mathcal C}$ is completely positive. We evaluate it on $\ket{0,0}$ in the block of $\ket w$ and on $\ket{1,1}$ in the block of $\ket{u_n}$. Its diagonal entries are $\bra{0,0}\widetilde{\mathcal C}(\ket w\!\bra w)\ket{0,0}$ and $p_{11}$. By \eqref{eq:branches} and the inequality of arithmetic and geometric means,
\begin{align*}
 |c|&\le\tfrac12\big(\sqrt{p_{00}p_{01}}+\sqrt{p_{10}p_{11}}\big)\\
 &\le\tfrac14\big(p_{00}+p_{01}+p_{10}+p_{11}\big)=\tfrac14\tr(\tau).
\end{align*}
So $\bra\Phi\tau\ket\Phi=\tfrac12(p_{00}+p_{11})+\operatorname{Re}c\le\tfrac34\tr(\tau)$. The output of the separable operation is a sum of such matrices $\tau$, and both sides of the inequality add up. So every two-qubit state that a separable operation produces from $\rho^{\otimes n}$ has fidelity at most $3/4$ with $\ket\Phi$.

\emph{Step 4: undistillability and separable outputs.} If $E_D(\rho)>0$, then for large $n$ an LOCC protocol turns $\rho^{\otimes n}$ into pairs of qubits close to copies of $\ket\Phi$. We keep one pair and discard the others. Every LOCC protocol is a separable operation~\cite{locc}, so this contradicts Step~3. Hence $E_D(\rho)=0$. Now suppose that a separable operation produces an entangled two-qubit state $\omega$ from $\rho^{\otimes n}$. We apply it to $m$ groups of $n$ copies and keep the result when all $m$ applications succeed. This is again a separable operation, and it produces $\omega^{\otimes m}$ with nonzero probability. Every entangled two-qubit state is distillable~\cite{horodecki97}, so for large $m$ an LOCC protocol turns $\omega^{\otimes m}$ into a pair of qubits close to $\ket\Phi$. The composition of separable operations is a separable operation, so this again contradicts Step~3. Hence every such two-qubit state is separable.

\emph{Step 5: the fidelity $1/2$.} For a unit product vector $\ket a\otimes\ket b$ of two qubits, $|\langle\Phi|a,b\rangle|^2=\tfrac12|a_0b_0+a_1b_1|^2\le\tfrac12$ by the Cauchy--Schwarz inequality. So every separable two-qubit state has fidelity at most $1/2$ with $\ket\Phi$~\cite{bdsw}. Every other Bell state is the image of $\ket\Phi$ under a unitary on Bob's qubit, so the same bound holds for it.
\end{proof}

The lemma also covers every state $\omega$ that Alice and Bob can prepare from $\rho$ by LOCC. Indeed, every LOCC operation is a separable operation~\cite{locc}. So a separable operation applied to copies of $\omega$, after the preparation, is a separable operation applied to copies of $\rho$, and the lemma applies to it.

\step{Secret key} The same inequality shows that the states have no distillable secret key either. In key distillation, Alice and Bob want equal and uniformly random bits about which an eavesdropper, Eve, knows nothing. Eve holds a purification of all the copies of $\rho$, and she keeps the whole public transcript of the protocol. An \emph{ideal secret bit} is a state $\tfrac12\sum_{i=0}^{1}\ket{i,i}\!\bra{i,i}\otimes\sigma_E$ of the two bits and of Eve's systems, in which Eve's state $\sigma_E$ does not depend on the bits. The \emph{distillable secret key} $K_D(\rho)$ is the best rate of such bits that two-way LOCC protocols reach on many copies of $\rho$, with an error that tends to zero. We use the precise definitions of Ref.~\cite[Definitions~4.1 and~5.1]{openai}.

\begin{proposition}[Zero distillable secret key]\label{prop:key}
Suppose that the reversed block matrix $\Omega_n(z)$ of \eqref{eq:reverse} is positive semidefinite for every $n$ and every $\ket z$. Then $K_D(\rho)=0$. Moreover, for every $n$, every two-way LOCC protocol on $\rho^{\otimes n}$ that ends with one bit for Alice and one bit for Bob has an output at trace distance at least $1/4$ from every ideal secret bit.
\end{proposition}

We follow the proof of Theorem~5.3 of Ref.~\cite{openai}, which is about states $(L\otimes M)(\ket\chi\!\bra\chi)$ for maps $L$ and $M$ that are completely positive and completely copositive. That proof uses this assumption only in the four-effect inequality, Lemma~3.2 of Ref.~\cite{openai}, applied to the copies of the state. We replace this inequality by the inequality \eqref{eq:four} below, which follows from the reversed block inequality.

\begin{proof}
\emph{Step 1: four positive observables.} We group Alice's $n$ output systems together, and Bob's as well. Let $X_0,X_1$ be positive semidefinite matrices on Alice's output, and let $X'_0,X'_1$ be positive semidefinite matrices on Bob's output. They need not form measurements. Let $D_{ij}=X_i\otimes X'_j$ and $p_{ij}=\tr(D_{ij}R_n)$. We show that
\begin{equation}\label{eq:four}
 \big\|\sqrt{D_{00}}\,R_n\sqrt{D_{11}}\big\|_1\le\tfrac14\sum_{i,j=0}^{1}p_{ij},
\end{equation}
where $\|\cdot\|_1$ is the trace norm. We repeat Steps~2 and~3 of the proof of Lemma~\ref{lem:fidelity}, with $\mathcal C_n$ in place of $\widetilde{\mathcal C}$ and with $E_i=(L^{\otimes n})^\dagger(X_i)$ and $F_j=(M^{\otimes n})^\dagger(X'_j)$. In Step~2 the left-hand side of \eqref{eq:pull} becomes $\tr\big(D_{ij}\,\mathcal C_n(\ket z\!\bra z)\big)$, and the inequalities \eqref{eq:branches} become
\begin{align*}
 \tr\big(D_{11}\,\mathcal C_n(\ket v\!\bra v)\big)&\le\tfrac14\,p_{01},\\
 \tr\big(D_{00}\,\mathcal C_n(\ket w\!\bra w)\big)&\le\tfrac14\,p_{10}.
\end{align*}
In Step~3 we multiply each of the two block matrices from both sides by the block diagonal matrix with the blocks $\sqrt{D_{00}}$ and $\sqrt{D_{11}}$. If a positive semidefinite block matrix has the diagonal blocks $A$ and $B$ and the off-diagonal block $C$, then $\|C\|_1\le\sqrt{\tr A\,\tr B}$~\cite[Lemma~3.1]{openai}. For the reversed block matrix $\Omega_n(v)$, this bounds the trace norm of $\sqrt{D_{00}}\,\mathcal C_n(\ket v\!\bra{u_n})\sqrt{D_{11}}$ by $\tfrac12\sqrt{p_{00}p_{01}}$. For the block matrix of $\mathcal C_n$ on $\ket w$ and $\ket{u_n}$, it bounds the trace norm of $\sqrt{D_{00}}\,\mathcal C_n(\ket w\!\bra{u_n})\sqrt{D_{11}}$ by $\tfrac12\sqrt{p_{10}p_{11}}$. The two matrices add up to $\sqrt{D_{00}}\,R_n\sqrt{D_{11}}$, because $\ket v+\ket w=\ket{u_n}$. The triangle inequality and the inequality of arithmetic and geometric means give \eqref{eq:four}. Both sides of \eqref{eq:four} are positively homogeneous in $R_n$, so the inequality also holds with $\rho^{\otimes n}$ in place of $R_n$.

\emph{Step 2: Eve's states.} Fix a protocol as in the proposition. A channel on Eve's systems can only decrease the trace distance, so we may give Eve the canonical purification of $\rho^{\otimes n}$~\cite[Lemma~4.3]{openai}. By Lemma~4.2 of Ref.~\cite{openai}, there are a measure $\mu$ on the transcripts and, for each transcript $t$, positive semidefinite matrices $X_i(t)$ on Alice's side and $X'_j(t)$ on Bob's side with the following property. In the branch in which Alice's bit is $i$ and Bob's bit is $j$, Eve's unnormalized state at the transcript $t$ has the density
\[
 \xi_{ij}(t)=\Big(\sqrt{\rho^{\otimes n}}\,D_{ij}(t)\sqrt{\rho^{\otimes n}}\Big)^{T}
\]
with respect to $\mu$, where $D_{ij}(t)=X_i(t)\otimes X'_j(t)$, and the integrals of the four traces add up to one. The root fidelity of $\xi_{00}(t)$ and $\xi_{11}(t)$ is the trace norm of $\sqrt{D_{00}(t)}\,\rho^{\otimes n}\sqrt{D_{11}(t)}$~\cite[Eq.~(5.5)]{openai}. So by \eqref{eq:four}, the integral of this root fidelity over all transcripts is at most $1/4$. Now let $\theta$ be the trace norm of the difference between the output of the protocol and an ideal secret bit. The same integral is at least $(1-\theta)/2$~\cite[Eq.~(5.9)]{openai}. So $\theta\ge1/2$, and the trace distance $\theta/2$ is at least $1/4$.

\emph{Step 3: zero key.} Suppose that $K_D(\rho)>0$. Then for large $n$ there are protocols that produce strings of key bits whose distance from an ideal key tends to zero. If Alice and Bob keep only the first bit of each string, the distance does not increase, and the result contradicts Step~2~\cite[proof of Theorem~5.3]{openai}. Hence $K_D(\rho)=0$.
\end{proof}

The conclusion $K_D(\omega)=0$ also holds if Alice and Bob can prepare $\omega$ from $\rho$ by LOCC, possibly by keeping an outcome with positive probability $q$. Consider a key protocol for $\omega$ that is secure when Eve holds a purification of the copies of $\omega$. It stays secure when Eve holds instead a purification of the copies of $\rho$ and the transcript of the preparation. Indeed, her systems then hold an extension of the copies of $\omega$, and every extension can be obtained from a purification by a channel on Eve's side. So a positive key rate from $\omega$ would give a positive key rate from $\rho$, at least $q$ times as large. Since $K_D(\rho)=0$, it follows that $K_D(\omega)=0$.

\subsection{Finding \texorpdfstring{$Q$}{Q} is a semidefinite program}

Fix the maps $L$ and $M$. Then $J$ and $R$ are fixed, and the conditions \eqref{eq:cert} are three linear matrix inequalities and one linear equation for the unknown matrix $Q$. So the set of matrices $Q$ that satisfy them is convex, and finding one is a semidefinite program. The next lemma describes all matrices that satisfy the last two conditions. For a basis vector $\ket p$ of the input, let
\[
 W_p=\mathcal C(\ket p\!\bra u),
\]
a matrix on the output, and let $N$ be the matrix with the blocks
\begin{equation}\label{eq:N}
 N_{pq}=W_q^{\dagger}\,R^{-1}\,W_p .
\end{equation}

\begin{lemma}[The form of $Q$]\label{lem:form}
A Hermitian matrix $Q$ satisfies \eqref{eq:cert-c} and \eqref{eq:cert-d} exactly when
\begin{equation}\label{eq:form}
 Q=N+Z^{\Gin}
\end{equation}
for a matrix $Z\succeq0$ with $ZV=0$.
\end{lemma}

So $Q=N$ is one choice, and every other choice adds the input transpose of a positive semidefinite matrix that vanishes on the range of $V$. Only the conditions \eqref{eq:cert-a} and \eqref{eq:cert-b} remain to be checked for a matrix of the form \eqref{eq:form}. We use two facts from the proof later: $V^\dagger JV=R$, and $NV=JV$.

\begin{proof}
We use two observations. First, let $X$ be Hermitian. The block of $XV$ with the index $p$ is $\sum_q u_qX_{pq}$, and the block of $X^{\Gin}V$ with the index $p$ is $\sum_q u_qX_{qp}$. Since $X_{qp}=X_{pq}^\dagger$ and the $u_q$ are real, the second block is the adjoint of the first. In particular, $XV=0$ implies $X^{\Gin}V=0$. Second, $\sum_q u_qW_q=\mathcal C(\ket u\!\bra u)=R$, and $R$ is Hermitian. In particular, $V^\dagger JV=\sum_pu_pW_p=R$.

By the second observation, $\sum_q u_qN_{pq}=R\,R^{-1}W_p=W_p$. This is also the block $\sum_q u_qJ_{pq}=\mathcal C(\ket p\!\bra u)$ of $JV$. So $(J-N)V=0$. The blocks of $N^{\Gin}$ are $W_p^\dagger R^{-1}W_q$, so $N^{\Gin}=\mathcal W R^{-1}\mathcal W^\dagger$ with $\mathcal W=\sum_p\ket p\otimes W_p^\dagger$, and $N^{\Gin}\succeq0$.

Now let $Q=N+Z^{\Gin}$ with $Z$ as in the lemma. Then $Q^{\Gin}=N^{\Gin}+Z\succeq0$. By the first observation, $Z^{\Gin}V=0$, so $(J-Q)V=0$.

Conversely, let $Q$ satisfy \eqref{eq:cert-c} and \eqref{eq:cert-d}. By \eqref{eq:cert-d} the blocks of $QV$ are the $W_p$. By the first observation, the blocks of $Q^{\Gin}V$ are the $W_p^\dagger$, so $Q^{\Gin}V=\mathcal W$ and $V^\dagger Q^{\Gin}V=\sum_pu_pW_p^\dagger=R$. For a positive semidefinite matrix $X$ and a matrix $V$ such that $V^\dagger XV$ is invertible, we have $X\succeq XV(V^\dagger XV)^{-1}V^\dagger X$. With $X=Q^{\Gin}$ the right-hand side is $\mathcal WR^{-1}\mathcal W^\dagger=N^{\Gin}$. So $Z=Q^{\Gin}-N^{\Gin}$ is positive semidefinite, and $ZV=\mathcal W-\mathcal WR^{-1}R=0$.
\end{proof}

\step{The program} The matrix $J-Q$ cannot be positive definite, because it vanishes on the range of $V$ by \eqref{eq:cert-d}. We take care of this range by adding $VV^\dagger$. Consider the semidefinite program in the variables $Q$, a Hermitian matrix, and $\delta$, a real number:
\begin{equation}\label{eq:sdp}
\begin{aligned}
 \text{maximize}\quad&\delta\\
 \text{subject to}\quad&(J-Q)\,V=0,\\
 &J-Q+VV^\dagger\succeq\delta I,\\
 &J+Q\succeq\delta I,\qquad Q^{\Gin}\succeq\delta I.
\end{aligned}
\end{equation}
A matrix $Q$ satisfies \eqref{eq:cert} exactly when $(Q,0)$ is a feasible point of \eqref{eq:sdp}. Indeed, let $(J-Q)V=0$. Since $J-Q$ is Hermitian, it maps the range of $V$ to zero and the orthogonal complement of this range into itself. The matrix $VV^\dagger$ is zero on the complement and positive definite on the range, because $V^\dagger V=\langle u|u\rangle\,I=r\,I$. So $J-Q+VV^\dagger\succeq0$ exactly when $J-Q\succeq0$. We call $Q$ a \emph{strict solution} if $(Q,\delta)$ is a feasible point of \eqref{eq:sdp} for some $\delta>0$. Then the conditions \eqref{eq:cert} hold with room to spare: $J-Q$ is positive definite on the orthogonal complement of the range of $V$, and $J+Q$ and $Q^{\Gin}$ are positive definite. A strict solution exists exactly when the optimal value of \eqref{eq:sdp} is positive.

By Lemma~\ref{lem:form} we can also take $Z$ as the variable, with $Q=N+Z^{\Gin}$. Then the equation $(J-Q)V=0$ holds whenever $ZV=0$, and it does not need to be imposed separately.

\begin{lemma}[Stability]\label{lem:stable}
Let $Q$ be a strict solution for the maps $L$ and $M$. Then there is an $\eta>0$ with the following property. If $L'$ and $M'$ are completely positive maps on the same input space whose product has a Choi matrix $J'$ with $\|J'-J\|_2\le\eta$, where $\|\cdot\|_2$ is the Hilbert--Schmidt norm, then $Q'=Q+J'-J$ is a strict solution for $L'$ and $M'$.
\end{lemma}

\begin{proof}
We have $J'-Q'=J-Q$, so the equation and the first inequality of \eqref{eq:sdp} are unchanged. Next, $J'+Q'=J+Q+2(J'-J)$ and $Q'^{\Gin}=Q^{\Gin}+(J'-J)^{\Gin}$. The input transpose does not change the Hilbert--Schmidt norm, and the operator norm is at most the Hilbert--Schmidt norm. So both matrices stay positive definite when $\eta$ is smaller than half the smallest eigenvalue of $J+Q$ and of $Q^{\Gin}$.
\end{proof}

So a strict solution stays a strict solution under a small change of the maps. If a family of maps crosses the PPT boundary within the neighborhood given by the lemma, the lemma gives solutions for the maps with NPT outputs there. This is why we look for maps whose outputs are close to the boundary.

\step{Symmetry} The matrices are large: $J$ has $r^2d^2$ rows. Our maps have symmetries, and these make the program much smaller. Suppose that unitaries $U_A,U_B$ on the input and $W_A,W_B$ on the output satisfy $(U_A\otimes U_B)\ket u=\ket u$, and that the unitary $g=\bar U_A\otimes\bar U_B\otimes W_A\otimes W_B$ commutes with $J$. Then $Q\mapsto gQg^\dagger$ maps solutions to solutions: it keeps $J\pm Q\succeq0$, it maps $V$ to $V(W_A\otimes W_B)$, and it transforms $Q^{\Gin}$ by the unitary $U_A\otimes U_B\otimes W_A\otimes W_B$. When the unitaries form a group, the average of a solution over the group is again a solution. So we may look for $Q$ among the matrices that commute with the group. This standard reduction~\cite{gatermann} leaves few variables. For example, the program in Sec.~\ref{sec:eight} has $24$ variables.

\subsection{Checking a solution}

A numerical solver returns an approximate solution. We turn it into an exact one, and then we check the exact one without using the solver. We assume that the maps have rational coefficients, so that $J$ and $N$ have rational entries. We round $Z=(Q-N)^{\Gin}$ to a Hermitian matrix with rational entries, and we project it onto the orthogonal complement of the range of $V$. As in the proof of Lemma~\ref{lem:form}, the new matrix $Q=N+Z^{\Gin}$ then satisfies $(J-Q)V=0$ exactly. If the solution is strict and the rounding is fine enough, the inequalities still hold. We then check the four conditions in exact rational arithmetic. Positivity is tested by Gaussian elimination over the integers~\cite{bareiss}. So no numerical tolerance is involved, and the check does not depend on the solver. The README of the accompanying program gives the details.

When the matrices commute with a group, positivity can be checked on smaller matrices. By Schur's lemma, a Hermitian matrix that commutes with a unitary representation of $U(d)$ is positive semidefinite exactly when its compression to the highest-weight vectors of each weight is positive semidefinite~\cite{fultonharris}. This turns a matrix with $r^2d^2$ rows into a list of blocks with few rows each.

\step{The accompanying program}\label{sec:code} The paper comes with a Python program \texttt{verify.py}, a README file and four JSON files. They can be downloaded from the arXiv source of this paper, in the folder \texttt{code}. Each JSON file is a certificate: it gives the input spaces, the local maps $L$ and $M$, the vector $\ket u$ and the matrix $Q$, and it names the Werner state that the maps produce. The files \texttt{d3.json}, \texttt{d5.json}, \texttt{d7.json} and \texttt{d8.json} are the certificates of Sec.~\ref{sec:computer}. Every number in a certificate is exact: an integer, or a fraction written as a string. The program uses only the standard library of Python. The README explains how to run it, how the JSON files are organized, and how to prepare a new certificate. For each certificate the program checks, in exact rational arithmetic,
\begin{enumerate}[(i)]
 \item that the local maps are completely positive, by testing their Choi matrices as in Fact~\ref{fact:choi}(\ref{fact:choi-b});
 \item that the output $R=V^\dagger JV$ is a positive multiple of $a\,I-b\,\swap$ with the ratio $t=b/a$ of the certificate, and that $1/d<t\le1/2$, so that the state is NPT by \eqref{eq:ratio};
 \item the four conditions \eqref{eq:cert}.
\end{enumerate}
If all checks pass, Lemmas~\ref{lem:reverse} and~\ref{lem:fidelity} show that the state is undistillable, and Lemma~\ref{lem:ops} extends this to the family of Theorem~\ref{thm:family}. The vector $\ket u$ may carry weights on the summands of the input. Section~\ref{sec:qutrits} explains why the lemmas still apply.

\step{Covariant maps} In the certificates, the input space of each party is a direct sum of subspaces $E$ of $(\mathbb C^d)^{\otimes k}$ that the operators $U^{\otimes k}$ map into themselves, such as the symmetric subspace. On each summand the local map is $X\mapsto\alpha\tr(X)\,I+\beta A(X)$ or $X\mapsto\alpha\tr(X)\,I+\beta A(X)^{T}$, where $A(X)$ is the partial trace of $X$ over all tensor factors but the first. Let $U_E$ be the restriction of $U^{\otimes k}$ to $E$, written in the real basis of $E$ that the certificate uses. For every matrix $X$ on $E$, a map $L$ of the first kind satisfies $L(U_EXU_E^\dagger)=U\,L(X)\,U^\dagger$, and a map of the second kind satisfies $L(\bar U_EX\bar U_E^\dagger)=U\,L(X)\,U^\dagger$. So $J$ commutes with an action of $U(d)$ in which a unitary acts on the outputs by $U\otimes U$, and on each input summand by $U^{\otimes k}$ or by its complex conjugate. This is a symmetry of the kind used above to reduce the program \eqref{eq:sdp}, and the certificates are chosen so that $Q$ commutes with this action too.

\step{Highest-weight vectors} So the program checks positivity on the highest-weight vectors, as above. For each pair of input summands, it computes the highest-weight vectors of their tensor product as the kernel of the raising operators. For $d\le8$ it checks their number against the Weyl character formula, and in every case it checks with Weyl's dimension formula that they account for the whole space. Then it takes the tensor product with the two outputs. Each output adds one box to a highest weight, so a weight $\mu$ becomes $\mu+e_a+e_b$ for two indices $a$ and $b$. The certificate gives $Q$ by its compressions to these vectors, in a basis that the README describes, and the program computes the compressions of $J$ from the maps. It checks $(J-Q)V=0$, and it checks $J\pm Q\succeq0$ block by block by exact Gaussian elimination. The matrix $Q$ acts only on the pairs of summands that the certificate lists. On the other pairs $J\pm Q=J$, which is positive semidefinite because the maps are completely positive.

\step{The input transpose} The program checks the third condition through the outputs. For every matrix, $Q^{\Gin}$ is the transpose of $Q^{\Gamma_{\rm out}}$, the partial transpose on the two outputs. A Hermitian matrix and its transpose have the same eigenvalues, so the program checks $Q^{\Gamma_{\rm out}}\succeq0$. This matrix commutes with the action in which the outputs carry $\bar U\otimes\bar U$. Its compressions to the highest-weight vectors of this action follow from the blocks of $Q$: for highest-weight vectors $v,v'$ of the same weight $\lambda$ and products $F_w,F_{w'}$ of lowering operators, $\langle F_wv,F_{w'}v'\rangle$ is $\langle v,v'\rangle$ times a number that depends only on $\lambda$, $w$ and $w'$. The program splits each highest-weight vector of the action with $\bar U\otimes\bar U$ on the outputs into its parts in the spaces $V_\mu\otimes\mathbb C^d\otimes\mathbb C^d$, where $V_\mu$ is an irreducible summand of the inputs. In each such space it projects onto the irreducible components of the original action, with $U\otimes U$ on the outputs. Their highest weights have the form $\mu+e_a+e_b$. By the same fact, the program can compute these projections inside $V_\mu\otimes\mathbb C^d\otimes\mathbb C^d$ from $\mu$ and from the norm of the highest-weight vector of $V_\mu$. It does not write out the vectors of $V_\mu$, so the check stays fast when the input summands are large.

\step{Mixtures and local maps} Tabia and Hsieh observed that the set of states with a solution of \eqref{eq:cert}, for some input spaces and some maps, contains the separable states and is closed under mixtures, tensor products and local completely positive maps~\cite{tabiahsieh}. We do not need this fact: in all our proofs it is enough that LOCC operations do not create distillable states.

\section{A proof without a computer for \texorpdfstring{$d\ge128$}{d >= 128}}
\label{sec:large}

In this section we prove Theorem~\ref{thm:examples}(a). Let $d\ge128$. We use the maps
\begin{equation}\label{eq:maps-large}
 L(X)=\tr(X)\,I-s\,X^{T},\qquad M(X)=\tr(X)\,I+b\,X ,
\end{equation}
on $d\times d$ matrices, with $s=1/4$ and $b>0$. The first term of each map replaces the system by noise, and the second term keeps it, transposed on Alice's side. The input space is $\mathcal H=\mathbb C^d$, and $\ket u=\ket{u_d}$.

\step{Step 1: the output} For basis vectors $\ket i$ and $\ket j$, let $\delta_{ij}=1$ if $i=j$ and $\delta_{ij}=0$ otherwise. Then $L(\ket i\!\bra j)=\delta_{ij}I-s\,\ket j\!\bra i$ and $M(\ket i\!\bra j)=\delta_{ij}I+b\,\ket i\!\bra j$. Since $\ket u\!\bra u=\sum_{i,j=1}^{d}\ket i\!\bra j\otimes\ket i\!\bra j$, we multiply and sum over $i,j=1,\dots,d$:
\begin{equation}\label{eq:R-large}
 R=\mathcal C(\ket u\!\bra u)=(d+b-s)\,I-sb\,\swap .
\end{equation}
By \eqref{eq:ratio}, the boundary of the PPT states is at $(d+b-s)/(sb)=d$, which gives $b=(4d-1)/(d-4)$. We lower the ratio $(d+b-s)/(sb)$ by one, from $d$ to $d-1$. This gives
\[
 b=\frac{4d-1}{d-5}=4+\frac{19}{d-5},\qquad R=sb\,\big((d-1)\,I-\swap\big).
\]
So $\rho=R/\tr(R)$ is the state of Theorem~\ref{thm:examples}(a), and $\rho^\Gamma\ket{u_d}=-\ket{u_d}/((d-1)d^2-d)$. For $d\ge128$ we have $4<b\le17/4$.

\step{Step 2: the choice of $Q$} We take $Q=N$, the matrix \eqref{eq:N}. In the program \eqref{eq:sdp} this is the point $Z=0$. By Lemma~\ref{lem:form}, $N$ satisfies \eqref{eq:cert-c} and \eqref{eq:cert-d}. It remains to show that $J-N\succeq0$ and $J+N\succeq0$.

\step{Step 3: notation} We number the four systems: $1$ is Alice's input, $2$ is Bob's input, $3$ is Alice's output and $4$ is Bob's output. Let $F_{ab}$ be the operator that exchanges the systems $a$ and $b$, and let $F=\swap$ on the output, so that $F_{34}=I\otimes F$. We use three maps from the output $\mathbb C^d\otimes\mathbb C^d$ into the space of the four systems,
\begin{align*}
 T_0\ket{x,y}&=\textstyle\sum_{k=1}^{d}\ket{k,k,x,y},\\
 T_1\ket{x,y}&=\textstyle\sum_{k=1}^{d}\ket{x,k,k,y},\\
 T_2\ket{x,y}&=\textstyle\sum_{k=1}^{d}\ket{y,k,x,k}.
\end{align*}
So $T_0=V$, $T_1=F_{13}T_0$ and $T_2=F_{14}T_0$. Counting the matching basis vectors gives
\begin{equation}\label{eq:gram-large}
 T_k^\dagger T_k=d\,I,\quad T_0^\dagger T_1=T_0^\dagger T_2=I,\quad T_1^\dagger T_2=F,
\end{equation}
and $F_{34}T_1=T_2F$. The Choi matrix of $L$ is $I-sF_{13}$, and the Choi matrix of $M$ is $I+b\,\ket u\!\bra u$ on the systems $2,4$, which is $I+b\,T_2T_2^\dagger$ on all four systems. The two factors act on different systems, so they commute, and $I-sF_{13}\succeq(1-s)\,I$. Hence
\begin{equation}\label{eq:Delta}
\begin{aligned}
 J&=(I-sF_{13})(I+b\,T_2T_2^\dagger)\succeq\Delta,\\
 \Delta&=(1-s)(I+b\,T_2T_2^\dagger).
\end{aligned}
\end{equation}

\step{Step 4: reduction to the complement of the range of $T_0$} Let $J_0=(JV)R^{-1}(JV)^\dagger$, which is positive semidefinite. We have $V^\dagger JV=R$ and $NV=JV$ by Sec.~\ref{sec:method}, and hence $J_0V=JV$. So the two Hermitian matrices $J-N$ and $J+N-2J_0$ vanish on the range of $V=T_0$, and they map its orthogonal complement into itself. In Steps~5 to~7 we show that for every vector $\ket\psi$ orthogonal to the range of $T_0$,
\begin{equation}\label{eq:two-bounds}
 \big|\bra\psi N\ket\psi\big|\le\tfrac23\bra\psi\Delta\ket\psi,\qquad \bra\psi J_0\ket\psi\le\tfrac16\bra\psi\Delta\ket\psi .
\end{equation}
These bounds imply the claim. Indeed, by \eqref{eq:Delta}, on the orthogonal complement $J-N\succeq\Delta-\tfrac23\Delta\succeq0$ and $J+N-2J_0\succeq(1-\tfrac23-\tfrac13)\Delta=0$. So $J-N\succeq0$ and $J+N-2J_0\succeq0$ everywhere, and $J+N\succeq0$ because $J_0\succeq0$.

\step{Step 5: the blocks $W_p$} Since $T_2T_2^\dagger\succeq0$, we have $\Delta\succeq(1-s)\,I$. By \eqref{eq:gram-large}, $T_2T_2^\dagger T_2=d\,T_2$, so
\begin{equation}\label{eq:Delta-T2}
 \Delta^{-1}T_2=\frac{T_2}{(1-s)\kappa},\qquad \kappa=1+bd .
\end{equation}

For the input basis vector $\ket p=\ket{i,j}$, the same computation as in Step~1 gives
\[
 W_{ij}=\delta_{ij}\,I+b\,E^{(4)}_{ji}(I-sF)-s\,E^{(3)}_{ji},
\]
where $E^{(3)}_{ji}=\ket j\!\bra i\otimes I$ and $E^{(4)}_{ji}=I\otimes\ket j\!\bra i$ act on the output. The matrix $JV$ has the blocks $W_p$. Collecting the three terms gives
\begin{equation}\label{eq:G}
 JV=T_0+G,\qquad G=b\,T_2(I-sF)-s\,T_1 .
\end{equation}

For a Hermitian matrix $H$ we write $\|H\|_\Delta=\|\Delta^{-1/2}H\Delta^{-1/2}\|$, where $\|\cdot\|$ is the operator norm. Then $|\bra\psi H\ket\psi|\le\|H\|_\Delta\bra\psi\Delta\ket\psi$ for every $\ket\psi$. We use the bounds
\begin{equation}\label{eq:T-norms}
 \|\Delta^{-1/2}T_1\|^2\le\frac d{1-s},\qquad \|\Delta^{-1/2}T_2\|^2=\frac d{(1-s)\kappa},
\end{equation}
which follow from $\Delta\succeq(1-s)\,I$, from \eqref{eq:Delta-T2} and from $T_k^\dagger T_k=d\,I$.

\step{Step 6: the bound on $J_0$} The output $R$ has the eigenvalues $c\pm\ell$, where $c=d+b-s$ and $\ell=sb$. Since $b(1-s)\ge s$, we have $c-\ell\ge d$, so $\|R^{-1}\|\le1/d$. Let $\ket\psi$ be orthogonal to the range of $T_0$. Then $T_0^\dagger\ket\psi=0$, and by \eqref{eq:G},
\[
 \bra\psi J_0\ket\psi=\bra\psi GR^{-1}G^\dagger\ket\psi\le\frac1d\,\|G^\dagger\Delta^{-1}G\|\,\bra\psi\Delta\ket\psi .
\]
By \eqref{eq:Delta-T2} and \eqref{eq:gram-large}, $T_2^\dagger\Delta^{-1}T_2=d\,I/((1-s)\kappa)$ and $T_2^\dagger\Delta^{-1}T_1=F/((1-s)\kappa)$. Also $T_1^\dagger\Delta^{-1}T_1\preceq d\,I/(1-s)$, $\|I-sF\|\le1+s$ and $\|(I-sF)F+F(I-sF)\|\le2(1+s)$. Since
\begin{align*}
 G^\dagger\Delta^{-1}G={}&\frac{b^2d}{(1-s)\kappa}\,(I-sF)^2+s^2\,T_1^\dagger\Delta^{-1}T_1\\
 &-\frac{sb}{(1-s)\kappa}\,\big[(I-sF)F+F(I-sF)\big] ,
\end{align*}
we get, with $w=b(1+s)$,
\[
 \frac1d\,\|G^\dagger\Delta^{-1}G\|\le\frac1{1-s}\Big(s^2+\frac{w^2+2sw/d}{\kappa}\Big).
\]
Now $s=1/4$, $w=5b/4\le85/16$, so $w^2+2sw/d<29$, and $\kappa>4d\ge512>22^2$. The right-hand side is at most $\tfrac43\big(\tfrac1{16}+\tfrac{29}{484}\big)<\tfrac16$. This proves the second bound in \eqref{eq:two-bounds}.

\step{Step 7: the bound on $N$} Let $\ket\psi$ be orthogonal to the range of $T_0$. We write $W_p=\delta_p\,I+G_p$, where $\delta_p=1$ if $p=(i,i)$ and $\delta_p=0$ otherwise, and $G_p$ is the block of $G$. In the expansion of $N_{pq}=W_q^\dagger R^{-1}W_p$, every term with a factor $\delta_p$ or $\delta_q$ gives a matrix of the form $T_0X$ or $XT_0^\dagger$, and its expectation value in $\ket\psi$ is zero. So $\bra\psi N\ket\psi=\bra\psi N_G\ket\psi$, where $N_G$ has the blocks $G_q^\dagger R^{-1}G_p$. We write $R^{-1}=x\,I+y\,F$, with
\[
 x=\frac c{c^2-\ell^2}\le\frac1d,\qquad y=\frac\ell{c^2-\ell^2}\le\frac{sb}{d^2}.
\]
Multiplying out the matrix units, and collecting terms with $F_{34}T_1=T_2F$ and $F_{34}F_{14}F_{34}=F_{13}$, gives
\begin{align*}
 N_G={}&b^2x\,F_{14}+s^2\big[(1+b^2)x+2by\big]F_{13}\\
 &-sb\,(bx+y)\,(F_{13}F_{14}+F_{14}F_{13})\\
 &+b^2y\,T_1FT_1^\dagger+s^2\big[(1+b^2)y+2bx\big]\,T_2FT_2^\dagger\\
 &-sb\,(x+by)\,(T_1T_2^\dagger+T_2T_1^\dagger).
\end{align*}
For example, the part $s^2x\,F_{13}$ comes from the blocks $E^{(3)}_{kl}E^{(3)}_{ji}=\delta_{jl}E^{(3)}_{ki}$, which are the blocks of $F_{13}$. We bound each term in $\|\cdot\|_\Delta$. A permutation of the systems has norm one, and $\|\Delta^{-1}\|\le1/(1-s)$. For the last three terms we use \eqref{eq:T-norms}. This gives
\begin{align*}
 (1-s)\,\|N_G\|_\Delta\le{}&b^2x+s^2\big[(1+b^2)x+2by\big]\\
 &+2sb\,(bx+y)+b^2yd\\
 &+s^2\big[(1+b^2)y+2bx\big]\frac d\kappa\\
 &+2sb\,(x+by)\frac d{\sqrt\kappa}.
\end{align*}
We insert $x\le1/d$, $y\le sb/d^2$, $b\le17/4$, $d\ge128$ and $\kappa>4d\ge512$, so that $\sqrt\kappa>22.6$. The six terms are at most $0.142$, $0.010$, $0.071$, $0.150$, $0.002$ and $0.098$. Their sum is less than $1/2$, so $\|N_G\|_\Delta<\tfrac12\cdot\tfrac43=\tfrac23$. This proves the first bound in \eqref{eq:two-bounds}.

\begin{proof}[Proof of Theorem~\ref{thm:examples}(a)]
The maps \eqref{eq:maps-large} are completely positive by Fact~\ref{fact:choi}(\ref{fact:choi-b}), because their Choi matrices $I-s\,\swap$ and $I+b\,\ket u\!\bra u$ are positive semidefinite. By Step~1, $\rho$ is NPT, hence entangled. By Steps~2 to~7, $Q=N$ satisfies \eqref{eq:cert}. Lemmas~\ref{lem:reverse} and~\ref{lem:fidelity} give the rest.
\end{proof}

\step{Dimensions three to five} With the input space $\mathbb C^d$, the maps \eqref{eq:maps-large} give no proof in the dimensions three to five, for any $s$ and $b$. Here $b>0$, and $0<s\le1$ because $L$ is completely positive only for $s\le1$. By \eqref{eq:ratio}, the output $(d+b-s)\,I-sb\,\swap$ of Step~1 is NPT exactly when $ds>1$ and
\[
 sb>\frac{s(d-s)}{ds-1}=1+\frac{1-s^2}{ds-1}.
\]
On the other hand, Lemma~\ref{lem:reverse} bounds $sb$ from above. Let $\ket z=\ket{1,1}$ and $\ket w=\ket u-\ket z$. We subtract the second block row and the second block column of $\Omega_1(z)$ from the first ones. This gives a positive semidefinite matrix with the blocks $\mathcal C(\ket w\!\bra w)$, $\mathcal C(\ket z\!\bra w)$, $\mathcal C(\ket w\!\bra z)$ and $\mathcal C(\ket z\!\bra z)$. Its entries at the output vector $\ket{2,1}$ in the first block and $\ket{1,2}$ in the second block form the matrix
\[
 \begin{pmatrix}d-1-s&-sb\\-sb&1-s\end{pmatrix}.
\]
So every matrix $Q$ for these maps requires $sb\le\sqrt{(1-s)(d-1-s)}$. For a fixed $s$, this upper bound grows with $d$, and the lower bound above decreases with $d$. So it is enough to show that the two bounds are incompatible for $d=5$ and $1/5<s\le1$. At $s=1$ the upper bound is zero. For $1/5<s<1$, the inequality of arithmetic and geometric means for $1-s$ and $1-s/4$ gives
\[
 \sqrt{(1-s)(4-s)}\le2-\frac{5s}{4}<\frac{s(5-s)}{5s-1}.
\]
The second inequality holds because the right side minus the middle term is $(21s^2-25s+8)/(4(5s-1))$, and the numerator has a negative discriminant. In the dimensions six and seven the two bounds are compatible, but we did not find a matrix $Q$ for these maps.

The estimates of Steps~6 and~7 are crude, and the matrix $N$ alone does not work in small dimensions. For $d=8$, $s=1/4$ and $b=31/3$, the value of $b$ in Step~1, a numerical computation gives the smallest eigenvalues $-12.16$ for $J-N$ and $-12.25$ for $J+N$. In the next section we use other maps and a matrix $Q$ with $Z\ne0$.

\section{A proof by hand in dimension eight}
\label{sec:eight}

In this section we prove Theorem~\ref{thm:examples}(b). We use maps of the form \eqref{eq:maps-large} with $s=3/5$, scaled to have integer coefficients, and with a noise parameter.

\step{Two local maps with a noise parameter} Let $d=8$ and $\ket u=\ket{u_8}$. For a number $a\ge0$ we define two maps on $8\times8$ matrices,
\begin{equation}\label{eq:maps}
\begin{aligned}
 L(X)&=5\tr(X)\,I-3X^{T},\\
 M_a(X)&=a\tr(X)\,I+2X ,
\end{aligned}
\end{equation}
and we let $\mathcal C_a=L\otimes M_a$. The first term of $M_a$ replaces Bob's system by noise, and the second term keeps it. So $a$ measures how much noise Bob adds.

\begin{lemma}\label{lem:maps8}
For every $a\ge0$,
\[
 \mathcal C_a(\ket{u}\!\bra{u})=(37a+10)\,I-6\,\swap .
\]
\end{lemma}

\begin{proof}
By \eqref{eq:maps},
\begin{align*}
 L(\ket i\!\bra j)&=5\,\delta_{ij}\,I-3\,\ket j\!\bra i,\\
 M_a(\ket i\!\bra j)&=a\,\delta_{ij}\,I+2\,\ket i\!\bra j .
\end{align*}
Since $\ket{u}\!\bra{u}=\sum_{i,j=1}^{8}\ket i\!\bra j\otimes\ket i\!\bra j$, we multiply the two right-hand sides and sum over $i$ and $j$:
\begin{align*}
 \sum_{i,j=1}^{8}5\,\delta_{ij}\,I\otimes a\,\delta_{ij}\,I&=40a\,I,\\
 \sum_{i,j=1}^{8}5\,\delta_{ij}\,I\otimes2\,\ket i\!\bra j&=10\,I,\\
 -\sum_{i,j=1}^{8}3\,\ket j\!\bra i\otimes a\,\delta_{ij}\,I&=-3a\,I,\\
 -\sum_{i,j=1}^{8}3\,\ket j\!\bra i\otimes2\,\ket i\!\bra j&=-6\,\swap .
\end{align*}
The sum of the four lines is $(37a+10)\,I-6\,\swap$.
\end{proof}

By Lemma~\ref{lem:maps8} and \eqref{eq:ratio}, the output is NPT exactly when $6/(37a+10)>1/8$, that is, when $a<38/37$. At $a=38/37$ the output is $48\,I-6\,\swap$, the boundary state. We lower the noise to $a=1$, so that the coefficient of $I$ drops by one, from $48$ to $47$. From now on
\[
 M=M_1,\qquad \mathcal C=L\otimes M,
\]
and Lemma~\ref{lem:maps8} gives
\begin{equation}\label{eq:R8}
 R=\mathcal C(\ket u\!\bra u)=47\,I-6\,\swap=2960\,\rho ,
\end{equation}
with $\rho=(47\,I-6\,\swap)/2960$. Recall that $\swap$ acts on $\mathbb C^8\otimes\mathbb C^8$, a space of dimension $64$. It has the eigenvalues $1$ and $-1$ and the trace $8$. So $47\,I-6\,\swap$ has the eigenvalues $41$ and $53$ and the trace $47\cdot64-6\cdot8=2960$, and $\rho$ is a state. As in Sec.~\ref{sec:higher}, $\rho^\Gamma\ket{u}=(47-48)\,\ket{u}/2960=-\ket u/2960$, and $\rho$ is NPT. For every $n$,
\[
 \rho^{\otimes n}=\frac{\mathcal C_n(\ket{u_n}\!\bra{u_n})}{2960^n},
\]
where $\mathcal C_n=(L\otimes M)^{\otimes n}$ and $\ket{u_n}=\ket u^{\otimes n}$.

\step{The maps are completely positive} We put the values of $L(\ket i\!\bra j)$ and $M(\ket i\!\bra j)=\delta_{ij}\,I+2\,\ket i\!\bra j$ from the proof of Lemma~\ref{lem:maps8} into the definition of the Choi matrix. The terms with $\delta_{ij}$ add up to a multiple of $\sum_i\ket i\!\bra i\otimes I=I$, and so
\begin{align*}
 J_L&=5\,I-3\sum_{i,j=1}^{8}\ket i\!\bra j\otimes\ket j\!\bra i=5\,I-3\,\swap,\\
 J_M&=I+2\sum_{i,j=1}^{8}\ket i\!\bra j\otimes\ket i\!\bra j=I+2\,\ket{u}\!\bra{u}.
\end{align*}
The two sums are the expansions of $\swap$ and of $\ket u\!\bra u=\swap^\Gamma$ from Sec.~\ref{sec:higher}. The matrix $\swap$ has the eigenvalues $1$ and $-1$, so $J_L$ has the eigenvalues $2$ and $8$. The matrix $\ket u\!\bra u$ has the eigenvalue $\langle u|u\rangle=8$ on $\ket u$ and $0$ on the orthogonal complement, so $J_M$ has the eigenvalues $17$ and $1$. Hence $L$ and $M$ are completely positive by Fact~\ref{fact:choi}(\ref{fact:choi-b}).

\step{Why $Q=J$ does not work} The map $\mathcal C$ is not completely copositive. Indeed, $(L\circ T)(X)=5\tr(X)\,I-3X$ and $(M\circ T)(X)=\tr(X)\,I+2X^{T}$, so as in the proof of Lemma~\ref{lem:maps8},
\begin{align*}
 J_{L\circ T}&=5\,I-3\,\ket{u}\!\bra{u},\\
 J_{M\circ T}&=I+2\,\swap .
\end{align*}
The first matrix has the eigenvalues $5$ and $5-3\cdot8=-19$, and the second has the eigenvalues $1+2=3$ and $1-2=-1$. By Fact~\ref{fact:choi}(\ref{fact:choi-d}), the Choi matrix of $\mathcal C_n\circ T$ is a tensor product of $n$ copies of each, after the systems are put in order, and the eigenvalues of a tensor product are the products of the eigenvalues of its factors. Taking $-19$ from one factor and a positive eigenvalue from each other factor gives the negative eigenvalue $(-19)\cdot3\cdot(5\cdot3)^{n-1}$. So $\mathcal C_n\circ T$ is not completely positive, for any $n$. Lemma~\ref{lem:reverse} is what replaces complete copositivity. The matrix $N$ alone does not work either: a numerical computation gives the smallest eigenvalues $-1.33$ for $J-N$ and $-2.11$ for $J+N$.

\step{The matrix $Q$} We add one term to $N$. We use the maps $T_0,T_1,T_2$ and the exchanges $F_{ab}$ of the four systems from Step~3 of Sec.~\ref{sec:large}, now with $d=8$. Recall that $T_0=V$ and $T_0^\dagger T_0=8\,I$. Let $S=F_{12}$ be the exchange of the two inputs, let $E=I-T_0T_0^\dagger/8$ be the projector onto the orthogonal complement of the range of $T_0$, and let $\Pi_-=(I-F)/2$ be the projector onto the antisymmetric subspace of the output. We set
\begin{equation}\label{eq:Q8}
 H=E\,T_2\,\Pi_-\,T_2^\dagger E,\qquad Q=N+(SHS)^{\Gin}.
\end{equation}
The matrix $Z=SHS$ is positive semidefinite, and $ZV=0$, because $ST_0=T_0$ and $ET_0=0$. So $Q$ satisfies \eqref{eq:cert-c} and \eqref{eq:cert-d} by Lemma~\ref{lem:form}. We found the maps and the term $H$ by a small computer search. The proof does not depend on how they were found.

\begin{lemma}\label{lem:q8}
The matrix $Q$ of \eqref{eq:Q8} satisfies the four conditions \eqref{eq:cert}. Moreover, $J+Q$ is positive definite.
\end{lemma}

We have just shown \eqref{eq:cert-c} and \eqref{eq:cert-d}. We prove $J-Q\succeq0$ and $J+Q\succ0$ by hand in Appendix~\ref{app:hand}. All matrices there have size at most three, and the last tests only compare integers.

\begin{proof}[Proof of Theorem~\ref{thm:examples}(b)]
The state $\rho$ is NPT by \eqref{eq:R8}, hence entangled. By Lemmas~\ref{lem:q8} and~\ref{lem:reverse}, the inequality \eqref{eq:reverse} holds for every $n$ and every $\ket z$. Lemma~\ref{lem:fidelity} gives $E_D(\rho)=0$ and the statement about separable operations.
\end{proof}

\step{The state of the first version} The first version of this paper proved that the state $(23\,I-3\,\swap)/1448$ is NPT and bound entangled. It used the map $L$ for Alice, the map $M'(X)=18\tr(X)\,I+37X$ for Bob, and a matrix $Q$ that came from a numerical search. Its $24$ coefficients are integers divided by $40$. This result also follows from the present paper, in two ways. First, the state has $p=1/181<1/29$, so it is covered by Theorem~\ref{thm:family}: Lemma~\ref{lem:ops} takes the state of Table~\ref{tab:intervals} with $d_0=7$ to it. It does not follow from Theorem~\ref{thm:examples}(b), because the state there has the smaller value $p=1/370$, and Lemma~\ref{lem:ops} only lowers $p$. Second, it has a proof by hand of the same kind as above. Let $\Pi_\pm=(I\pm F)/2$. With the maps $L$ and $M'$, we define $Q$ by
\[
 S\,Q^{\Gin}S=\sum_{m,n=0}^{2}T_m\big(k^+_{mn}\Pi_++k^-_{mn}\Pi_-\big)T_n^\dagger ,
\]
where
\[
 4k^+=\begin{pmatrix}44&-81&89\\-81&150&-162\\89&-162&190\end{pmatrix},\quad
 2k^-=\begin{pmatrix}17&12&32\\12&18&0\\32&0&114\end{pmatrix}.
\]
The leading principal minors of $4k^+$ are $44$, $39$ and $240$, and those of $2k^-$ are $17$, $162$ and $36$. So $k^+$ and $k^-$ are positive definite, and $Q^{\Gin}\succeq0$. The other three conditions follow by the steps of Appendix~\ref{app:hand}, with matrices of size at most three. We checked them in exact arithmetic.

\section{Computer-aided proofs}
\label{sec:computer}

Below dimension eight we did not find a matrix $Q$ for maps with the input space $\mathbb C^d$. The states in this section use larger input spaces. One of them is a state in dimension eight, which lies farther from the boundary than the state of Sec.~\ref{sec:eight}. Their matrices $Q$ are too large to check by hand, and we check them with programs in exact arithmetic, as described in Sec.~\ref{sec:method}.

\subsection{Two qutrits}
\label{sec:qutrits}

Brand\~ao found a certificate for the state $(20000\,I-6877\,\swap)/159369$~\cite{brandao}. While found independently, its local maps extend those of Tabia and Hsieh, who gave an example of an NPT bound entangled two-qutrit Werner state~\cite{tabiahsieh}. We extended his certificate: we changed the parameters of his maps, added a fourth summand to his input space, and solved the program \eqref{eq:sdp} again. The new certificate proves that a state slightly farther from the boundary is undistillable. This subsection describes it.

\step{The input space} For $k\ge1$, let $\mathcal S_k$ be the symmetric subspace of $(\mathbb C^3)^{\otimes k}$, the space of $k$ bosons in three modes, of dimension $(k+1)(k+2)/2$. For a matrix $X$ on $\mathcal S_k$, let $A(X)=\tr_{2,\dots,k}(X)$ be its one-particle reduced matrix. Each party holds four such spaces, with $k_0=80$, $k_1=17$, $k_2=8$ and $k_3=6$, and a flag qubit for each of them. Brand\~ao's certificate uses the first three. So the input space of each party is $\bigoplus_{i=0}^{3}\mathbb C^2\otimes\mathcal S_{k_i}$, of dimension $7130$. For the summand $i$ let
\begin{align*}
 M_i(X)&=b_i\tr(X)\,I+A(X),\\
 L_i(X)&=g_i\big(a_i\tr(X)\,I-A(X)^{T}\big),
\end{align*}
with the numbers of Table~\ref{tab:qutrit}. If the flag of the summand $i$ is $\ket0$, Alice applies $M_i$ and Bob applies $L_i$. If it is $\ket1$, they exchange their roles. The maps ignore the blocks between different summands and between different flag values. The vector $\ket u$ is
\[
 \ket u=\sum_{i=0}^{3}c_i\,\big(\ket{i0,i0}+\ket{i1,i1}\big)\otimes\ket{u_{k_i}},
\]
where $\ket{u_k}=\sum_a\ket{a,a}$ is the maximally entangled vector of $\mathcal S_k\otimes\mathcal S_k$ for the orthonormal basis of normalized symmetrized standard basis vectors $\ket a$. Like $\ket{u_d}$, it is not normalized. Here $c_0^2=376/1107$ and $c_1^2=c_2^2=c_3^2=1/16$. Then $R$ is a positive multiple of $a\,I-b\,\swap$, where $a$ and $b$ are integers with $48$ and $47$ digits, and $t=b/a\approx0.3438886631$. We call the normalized state the certified state. It has $p\approx0.0119219$.

\begin{table}[t]
\caption{\label{tab:qutrit}The parameters of the maps $M_i$ and $L_i$ in Sec.~\ref{sec:qutrits}. The upper three rows are Brand\~ao's values~\cite{brandao}. The lower four rows are the exact values in our certificate.}
\begin{ruledtabular}
\begin{tabular}{rlll}
$k_i$ & $a_i$ & $b_i$ & $g_i$\\
\hline
$80$ & $\frac{8021}{8000}$ & $\frac{3428010459039151}{469477316750000000}$ & $1$\\[2pt]
$17$ & $\frac{173}{170}$ & $\frac{418}{15625}$ & $\frac{34818916}{890625}$\\[2pt]
$8$ & $\frac{83}{80}$ & $\frac{54309}{1000000}$ & $\frac{4439056}{78125}$\\
\hline
$80$ & $1.00302570155824$ & $0.00730175950304$ & $1.00070669805628$\\
$17$ & $1.0168443940293$ & $0.02669554060611$ & $41.87776652675296$\\
$8$ & $1.03890256539105$ & $0.05425867128999$ & $25.73459110294401$\\
$6$ & $1.05$ & $0.07$ & $32.23128951680229$\\
\end{tabular}
\end{ruledtabular}
\end{table}

\step{The matrix $Q$} The matrix $Q$ commutes with an action of $U(3)$. It vanishes on the inputs in which the flags of Alice and Bob differ, and it connects the flag values $00$ and $11$ of each summand, as well as different summands. Brand\~ao found his matrix $Q$ by solving the program \eqref{eq:sdp} with this symmetry, and he gave it with rational entries on the highest-weight vectors of each weight. For our certificate, we changed the parameters of the first three summands, added the fourth summand, and solved the program again numerically. We combined two numerical solutions: one with the largest value of $t$, and one with a positive margin $\delta$, which leaves room for rounding. Then we rounded the combination to rational numbers and corrected it so that the equation $(J-Q)V=0$ holds exactly. We wrote the result in the JSON format of the accompanying program (Sec.~\ref{sec:code}).

\step{The check} The matrices have $7130^2\cdot9$ rows. The accompanying program reduces them to the highest-weight vectors of the action of $U(3)$ (Sec.~\ref{sec:code}). There are $139$ blocks of $J-Q$ and $J+Q$, with at most $32$ rows each, and $145$ blocks of $Q^{\Gin}$. The program checks the equation $(J-Q)V=0$ and the positivity of all these blocks, and it recovers the certified state exactly. On a laptop the check takes about eight minutes.

The vector $\ket u$ here is not the vector $\sum_a\ket{a,a}$ of Sec.~\ref{sec:method}, because of the weights $c_i$. It is $(K\otimes I)\sum_a\ket{a,a}$, where $K$ is the positive diagonal matrix that is $c_i$ times the identity on the summand $i$ of Alice's input. We replace each of Alice's maps $\Lambda$ by $X\mapsto\Lambda(KXK)$. Then $\ket u$ becomes $\sum_a\ket{a,a}$, the output $R$ does not change, and $J$ and $Q$ change by the same congruence with $K$, which keeps the four conditions. So the certified state is undistillable by Lemmas~\ref{lem:reverse} and~\ref{lem:fidelity}.

\step{The data file} The file \texttt{d3.json} of the accompanying program describes the same certificate with another scaling of the summand $i=0$. In the file, the coefficient of this summand in $\ket u$ is $c_0^2$ instead of $c_0$, and Alice's maps $M_0$ and $L_0$ are multiplied by $1/c_0^2$. Table~\ref{tab:qutrit} lists the maps without this factor. The output $R$ is the same. The congruence on Alice's input with the diagonal matrix that is $1/c_0$ on the summand $i=0$ and the identity on the other summands takes the matrices $J$ and $Q$ of this section to those of the file.

\step{The state $\rho_3$} Theorem~\ref{thm:examples}(c) is about a simpler state, slightly closer to the boundary than the certified state:
\[
 \rho_3=\frac{1579\,I-543\,\swap}{12582}.
\]
As in Sec.~\ref{sec:higher}, its partial transpose has the eigenvalue $(1579-1629)/12582=-25/6291$ on $\ket{u_3}$, and $p(\rho_3)=25/2097\approx0.0119218$. This value is less than $p$ of the certified state, so $\rho_3$ is undistillable by Lemma~\ref{lem:ops}. This proves Theorem~\ref{thm:examples}(c).

\step{Earlier two-qutrit states} Tabia and Hsieh gave the example $(187499\,I-62500\,\swap)/1499991$~\cite{tabiahsieh}. Their input space is $\mathbb C^2\otimes\mathcal S_{12}$, with the maps $\tr(X)\,I+c\,A(X)$ and $\tr(X)\,I+b\,A(X)^{T}$ and the same rule for the flag, and they gave $Q$ by the rational coefficients of $398$ pairing diagrams. We changed $b$ and $c$, and this gave the states $((3m-1)\,I-m\,\swap)/(24m-9)$ for $m=32000$, $18000$ and $10500$. For $m=32000$ we kept their matrix $J-Q$, as in the proof of Lemma~\ref{lem:stable}. For $m=18000$ we multiplied $J-Q$ by $19301/19300$. For $m=10500$ we multiplied it by $10786/10785$ and subtracted $99/10^6$ times the projector onto the orthogonal complement of the range of $V$. We use the last state in Sec.~\ref{sec:279}. Brand\~ao's certificate is for $(20000\,I-6877\,\swap)/159369$, with $p=631/53123$. His letter also contains a certificate with the input space $\mathbb C^2\otimes\mathcal S_{18}$ for $(11249\,I-3750\,\swap)/89991$, and the state $(32\,I-11\,\swap)/255$, which he obtains from $(20000\,I-6877\,\swap)/159369$ by mixing. He also gives a perturbation argument that reaches slightly beyond $(20000\,I-6877\,\swap)/159369$, to $t\approx0.3438500002$~\cite{brandao}. Before we added the fourth summand, we had kept his input space and changed the parameters of his maps and his matrix $Q$. This gave $(500000\,I-171931\,\swap)/3984207$. Our own solution of the program on symmetric subspaces gave $(47\,I-16\,\swap)/375$. All these states have smaller values of $p$ than $\rho_3$, so they follow from $\rho_3$ by Lemma~\ref{lem:ops}, and we do not present their certificates. Tabia and Hsieh also gave Werner states in the local dimensions four to seven, $((dm-1)\,I-m\,\swap)/(d^2(dm-1)-dm)$ with $m=8500,62,45,37$. Their values $p=1/127496$, $1/1483$, $1/1569$ and $1/1769$ lie below $p_0(d)$, so Theorem~\ref{thm:family} covers them as well.

\subsection{Dimensions five, seven and eight}
\label{sec:wide}

The states in Table~\ref{tab:intervals} for $d_0=5,7,8$ come from the same method with other input spaces. In each case the input space is an irreducible representation of $U(d_0)$, given by a Young diagram $\lambda$ (Table~\ref{tab:wide}), and the local maps are covariant completely positive maps (Sec.~\ref{sec:code}). For $d_0=7$ the input space is the antisymmetric subspace of $(\mathbb C^7)^{\otimes3}$, and the maps are
\begin{align*}
 L(X)&=37\tr(X)\,I-63\,A(X)^{T},\\
 M(X)&=37\tr(X)\,I+147\,A(X),
\end{align*}
where $A(X)$ is again the one-particle reduced matrix. For $d_0=8$ the input space is the antisymmetric subspace of $(\mathbb C^8)^{\otimes4}$, and the maps are
\begin{align*}
 L(X)&=9817\tr(X)\,I-22361\,A(X)^{T},\\
 M(X)&=62676209\tr(X)\,I+331582300\,A(X).
\end{align*}

For each state we found $Q$ by the program \eqref{eq:sdp} with symmetry, and we rounded it as in Sec.~\ref{sec:method}. The accompanying program checks the four conditions in exact arithmetic on the blocks listed in Table~\ref{tab:wide}. On a laptop each check takes less than a minute for $d_0=5,7$ and about five minutes for $d_0=8$. So the states of Table~\ref{tab:wide} are undistillable by Lemmas~\ref{lem:reverse} and~\ref{lem:fidelity}. For $d_0=5,7$, $Q$ is a strict solution. In the exact check every pivot is positive, except for the two zero pivots of $J-Q$ that the equation $(J-Q)V=0$ forces. The Choi matrices of the local maps are positive definite as well. For $d_0=8$ we have not checked whether $Q$ is strict.

We found further states with the same method: $(23\,I-6\,\swap)/344$ in dimension four, with the input space of the Young diagram $(4,4)$, and $(23\,I-4\,\swap)/804$ and $(35\,I-6\,\swap)/1224$ in dimension six, with the input space $\wedge^2\mathbb C^6$. Their values $p=1/86$, $1/134$ and $1/204$ lie below $p_0(d)$, so Theorem~\ref{thm:family} covers them, and we do not present their certificates. The same holds for the state of Theorem~\ref{thm:examples}(b) and for the states of Theorem~\ref{thm:examples}(a), whose values of $p$ are at most $1/370$. We present them because their proofs need no computer.

\begin{table}[t]
\caption{\label{tab:wide}The states of Table~\ref{tab:intervals} in the local dimensions $d_0=5,7,8$. The input space is the irreducible representation of $U(d_0)$ with the Young diagram $\lambda$, of dimension $r$. The last column gives the number of blocks in the exact check and the largest number of rows of a block.}
\begin{ruledtabular}
\begin{tabular}{llrlc}
$d_0$ & $\lambda$ & $r$ & State & Blocks\\
\hline
$5$ & $(2,2)$ & $50$ & $(23\,I-5\,\swap)/550$ & $20$, $\le8$\\
$7$ & $(1,1,1)$ & $35$ & $(17\,I-3\,\swap)/812$ & $12$, $\le4$\\
$8$ & $(1,1,1,1)$ & $70$ & $(25000\,I-4427\,\swap)/1564584$ & $15$, $\le4$\\
\end{tabular}
\end{ruledtabular}
\end{table}

\subsection{Proof of Theorem~\ref{thm:family}}

The state $\rho_3$ of Sec.~\ref{sec:qutrits} is NPT and undistillable, with $p(\rho_3)=25/2097$. Lemma~\ref{lem:ops} with $\tau=\rho_3$ gives the theorem for $d\ge3$ with this value of $p_0$. In the same way, the states of Table~\ref{tab:wide} give the theorem with $p_0=1/55$ for $d\ge5$, with $p_0=1/29$ for $d\ge7$ and with $p_0=496/9313$ for $d\ge8$, and Table~\ref{tab:intervals} lists for each $d$ the largest value that applies. The values of $p$ of these states follow from $\tr\big(\swap\,(a\,I-b\,\swap)\big)=d\,(a-bd)$. The statement about separable operations follows from Lemma~\ref{lem:fidelity} and the remark after it, because Lemma~\ref{lem:ops} prepares $\omega$ by LOCC from a state to which Lemma~\ref{lem:fidelity} applies.

The values $p_0(d)$ are lower bounds on the thresholds $p_*(d)$ of Theorem~\ref{thm:threshold}, and for $d\le7$ none of them is sharp. For $d=3,4$, the certified state of Sec.~\ref{sec:qutrits} has a larger value of $p$ than $\rho_3$. For $d\ge5$ we use that the solutions of Table~\ref{tab:wide} are strict. In dimension seven, we lower the coefficient $37$ of $\tr(X)\,I$ in $M$ by a small amount. The map stays completely positive, because its Choi matrix is positive definite. The output moves further beyond the boundary, and by Lemma~\ref{lem:stable} the four conditions still hold for a nearby matrix $Q$. So $p_*(7)>1/29$. The same argument in dimension five gives $p_*(5)>1/55$. By Theorem~\ref{thm:threshold}(ii), $p_*(d)>p_0(d)$ for $3\le d\le7$. For $d\ge8$ we know only that $p_*(d)\ge496/9313$, because we have not checked whether the solution in dimension eight is strict. So $c\ge496/9313$.

\section{Beyond Werner states}
\label{sec:nonwerner}

In this section we use only that the states of Theorems~\ref{thm:family} and~\ref{thm:examples} are undistillable. For $d\ge3$, let $\omega_d$ be the Werner state with $p(\omega_d)=p_0(d)$, and write it as $\omega_d=a\,I-b\,\swap$. It is NPT and undistillable by Theorem~\ref{thm:family}.

\subsection{Mixtures with separable states}

\begin{theorem}[Mixtures with separable states]\label{thm:mixture}
Let $\sigma$ be a separable state on $\mathbb C^d\otimes\mathbb C^d$, let $0<\lambda\le1$, and let
\[
 \tau=\lambda\,\omega_d+(1-\lambda)\,\sigma,\qquad G=\lambda a\,I+(1-\lambda)\,\sigma^\Gamma .
\]
The state $\tau$ is undistillable. It is NPT, and hence bound entangled, exactly when
\[
 \lambda b\,\bra{u_d}G^{-1}\ket{u_d}>1 .
\]
\end{theorem}

\begin{proof}
Alice and Bob can prepare $\tau$ from $\omega_d$ by LOCC: with probability $1-\lambda$ they replace $\omega_d$ by $\sigma$. So $\tau$ is undistillable. Its partial transpose is $\tau^\Gamma=G-\lambda b\,\ket{u_d}\!\bra{u_d}$. The matrix $G$ is positive definite, because $\sigma^\Gamma\succeq0$ for a separable state. So $\tau^\Gamma$ has a negative eigenvalue exactly when $G^{-1/2}\tau^\Gamma G^{-1/2}$ has one. This matrix is the identity minus a positive semidefinite matrix of rank one, and its smallest eigenvalue is $1-\lambda b\,\bra{u_d}G^{-1}\ket{u_d}$.
\end{proof}

For example, let $\tau=\tfrac12\rho+\tfrac12\ket{1,2}\!\bra{1,2}$ with the state $\rho$ of Theorem~\ref{thm:examples}(b). The same argument applies to it. The product state is equal to its partial transpose, and it is orthogonal to $\ket{u_8}$. So $\tau^\Gamma\ket{u_8}=-\ket{u_8}/5920$, and $\tau$ is NPT and bound entangled. Its two local states are not maximally mixed, so $\tau$ is not a Werner state.

\subsection{A ball with one separable state}

\begin{theorem}[A ball with one separable state]\label{thm:ball}
Let $d\ge3$, let $e=dp_0(d)/(2+dp_0(d))$ and $\eta=(1-e)\,\omega_d+e\,I/d^2$. Consider the Hermitian matrices of trace one on $\mathbb C^d\otimes\mathbb C^d$ with the Hilbert--Schmidt distance. The closed ball around $\eta$ with the radius
\[
 r_d=\frac{e}{d\sqrt{d^2-1}}
\]
consists of states. One of them is the separable state $\pi_d$ of Lemma~\ref{lem:ops}. All the others are NPT and bound entangled.
\end{theorem}

\begin{proof}
Every Hermitian matrix of trace one whose Hilbert--Schmidt distance from $I/d^2$ is at most $1/(d\sqrt{d^2-1})$ is a state, and Gurvits and Barnum proved that it is separable~\cite[Corollary~3]{gurvits}. The map $\sigma\mapsto(1-e)\,\omega_d+e\,\sigma$ sends this ball onto the ball of the theorem. So every point of our ball is a mixture of $\omega_d$ with a separable state, and it is undistillable as in Theorem~\ref{thm:mixture}.

We have $p(I/d^2)=-1/d$, so $p(\eta)=(1-e)\,p_0(d)-e/d=e/d$ by the choice of $e$. On the matrices of trace one, $p(\tau)=-\tr\big((\swap-I/d)\,\tau\big)-1/d$, and the matrix $\swap-I/d$ has trace zero and Hilbert--Schmidt norm $\sqrt{d^2-1}$. So the distance from $\eta$ to the hyperplane $p=0$ is
\[
 \frac{p(\eta)}{\sqrt{d^2-1}}=\frac{e}{d\sqrt{d^2-1}}=r_d .
\]
The ball touches the hyperplane in one point, which lies on the line through $\eta$ in the direction $\swap-I/d$. This is the line of the Werner states, so the point is $\pi_d$. Every other point $\tau$ of the ball has $p(\tau)>0$, and it is NPT by \eqref{eq:witness}.
\end{proof}

For two qutrits, $e=25/1423$. The center of the ball is the Werner state
\[
 \eta=\frac{1604\,I-543\,\swap}{12807},
\]
and the radius is $r_3=25/(4269\sqrt8)$. The value of $e$ is the largest one for which the ball lies in the half-space $p\ge0$. Tabia and Hsieh used the same mixture with a small $e$ to obtain open sets of NPT bound entangled states~\cite{tabiahsieh}.

\subsection{The conjecture of Bandyopadhyay and Roychowdhury}

Most members of the family \eqref{eq:br} are not Werner states. The vector $\ket\varphi$ need not be maximally entangled, $\sigma$ can be any state with $\bra\varphi\sigma^\Gamma\ket\varphi<0$, and the two local dimensions may differ. The Werner states form a special case, which we treat first.

We return to the states $\rho_\epsilon$ of Eq.~\eqref{eq:br}. Let
\[
 \pi_\varphi=\frac{(I-\ket\varphi\!\bra\varphi)^{\Gamma}}{D-1},
\]
so that $\rho_\epsilon=\epsilon\,\sigma+(1-\epsilon)\,\pi_\varphi$. Since $\bra\varphi(I-\ket\varphi\!\bra\varphi)\ket\varphi=0$, we have $\bra\varphi\rho_\epsilon^\Gamma\ket\varphi=\epsilon\,\bra\varphi\sigma^\Gamma\ket\varphi$. So $\rho_\epsilon$ is NPT for $\epsilon>0$.

\step{Werner members} After the first version of this paper appeared, Bandyopadhyay pointed out that its state belongs to this family~\cite{bandyopadhyay-note}. Take $d_A=d_B=d$, $\ket\varphi=\ket{u_d}/\sqrt d$, and let $\sigma=A_d=(I-\swap)/(d(d-1))$ be the normalized projector onto the antisymmetric subspace. This is the choice~(12) of Ref.~\cite{br03} with equal weights. Since $(\ket{u_d}\!\bra{u_d})^\Gamma=\swap$, we have $\pi_\varphi=(d\,I-\swap)/(d(d^2-1))=\pi_d$, the separable Werner state of Lemma~\ref{lem:ops}. So
\begin{equation}\label{eq:br-werner}
 \rho_\epsilon=(1-\epsilon)\,\pi_d+\epsilon\,A_d=\frac{(d+\epsilon)\,I-(1+d\epsilon)\,\swap}{d\,(d^2-1)} .
\end{equation}
We have $p(\pi_d)=0$ and $p(A_d)=1$, so $p(\rho_\epsilon)=\epsilon$. For this choice the family is the part of the line of Werner states with $0\le p\le1$, and $\epsilon$ is our parameter $p$. For $d=8$ and $\epsilon=1/370$ it is the state of Theorem~\ref{thm:examples}(b). By Theorem~\ref{thm:threshold}, $\rho_\epsilon$ is bound entangled exactly when $0<\epsilon\le p_*(d)$, and by Theorem~\ref{thm:family} this holds for $0<\epsilon\le p_0(d)$. For large $\epsilon$ the state can be distilled. For example, $A_d$ is distillable from one copy.

\step{All members} As Bandyopadhyay and Roychowdhury observed, the state $\pi_\varphi$ lies on the surface of the largest separable ball around $I/D$~\cite{br03}. Our proof of Theorem~\ref{thm:br} uses this fact. We write $\rho_\epsilon$ as a mixture of a filtered NPT bound entangled state of two qutrits and a remainder that lies in the separable ball.

\begin{proof}[Proof of Theorem~\ref{thm:br}]
Write $f(X)=\bra\varphi X^\Gamma\ket\varphi$ for Hermitian matrices $X$, and let $a=-f(\sigma)>0$. Then $f(\pi_\varphi)=0$.

\emph{Step 1: the separable ball.} Every Hermitian matrix of trace one whose Hilbert--Schmidt distance from $I/D$ is at most $1/\sqrt{D(D-1)}$ is a separable state~\cite[Corollary~3]{gurvits}. The partial transpose does not change the Hilbert--Schmidt norm, and $\pi_\varphi-I/D=\big(I/D-\ket\varphi\!\bra\varphi\big)^\Gamma/(D-1)$. So $\|\pi_\varphi-I/D\|_2^2=1/(D(D-1))$, and $\pi_\varphi$ lies on the boundary of the ball. For a Hermitian matrix $H$ of trace zero, $\tr\big((\pi_\varphi-I/D)H\big)=-f(H)/(D-1)$, so
\begin{equation}\label{eq:ball-expansion}
 \big\|\pi_\varphi+tH-I/D\big\|_2^2=\frac1{D(D-1)}-\frac{2t\,f(H)}{D-1}+t^2\|H\|_2^2 .
\end{equation}
Here $\|\cdot\|_2$ is the Hilbert--Schmidt norm. In particular, if $f(H)>0$, then $\pi_\varphi+tH$ is separable for $0\le t\le2f(H)/((D-1)\|H\|_2^2)$.

\emph{Step 2: a filtered qutrit state.} Let $\rho_3=(1579\,I-543\,\swap)/12582$ be the state of Theorem~\ref{thm:examples}(c). Write $\ket\varphi=\sum_is_i\ket{\alpha_i}\otimes\ket{\beta_i}$ in Schmidt form, with $s_1,s_2,s_3>0$. Define the maps $K_A$ and $K_B$ from $\mathbb C^3$ to Alice's and Bob's spaces by $K_A\ket i=s_i^{-1}\ket{\alpha_i}$ and $K_B\ket i=\ket{\bar\beta_i}$ for $i=1,2,3$, where the bar is the complex conjugate in the basis used for $\Gamma$. Then $(K_A^\dagger\otimes K_B^{T})\ket\varphi=\ket{u_3}$. Let $\omega$ be the state $(K_A\otimes K_B)\rho_3(K_A\otimes K_B)^\dagger$ after normalization. Since $\big((K_A\otimes K_B)X(K_A\otimes K_B)^\dagger\big)^\Gamma=(K_A\otimes\bar K_B)X^\Gamma(K_A\otimes\bar K_B)^\dagger$, the number $f(\omega)$ is a positive multiple of $\bra{u_3}\rho_3^\Gamma\ket{u_3}=-p(\rho_3)<0$. Let $a'=-f(\omega)>0$.

\emph{Step 3: the decomposition.} Let $\gamma=2a/a'$ and $H=\sigma-\pi_\varphi-\gamma(\omega-\pi_\varphi)$. Then $H$ has trace zero and $f(H)=-a+2a=a$. For $\gamma\epsilon<1$ we have
\begin{align*}
 \rho_\epsilon&=\gamma\epsilon\,\omega+(1-\gamma\epsilon)\,\tau_\epsilon,\\
 \tau_\epsilon&=\pi_\varphi+\frac{\epsilon}{1-\gamma\epsilon}\,H .
\end{align*}
By Step~1, $\tau_\epsilon$ is a separable state whenever $0<\epsilon\le\epsilon_\infty$, where
\[
 \epsilon_\infty=\min\Big\{1,\ \frac{2a}{(D-1)\|H\|_2^2+2a\gamma}\Big\}.
\]
The cap at $1$ is the end of the family~\eqref{eq:br}. The second entry of the minimum is less than $1/\gamma$, so $\gamma\epsilon_\infty<1$.

\emph{Step 4: undistillability.} Alice and Bob can prepare $\rho_\epsilon$ from $\rho_3$ by LOCC. They apply the local filters $K_A$ and $K_B$, rescaled so that they become Kraus operators of local measurements, and keep the state when both filters succeed. This gives $\omega$. Then, with probability $1-\gamma\epsilon$, they replace $\omega$ by the separable state $\tau_\epsilon$. The state $\rho_3$ is undistillable by Theorem~\ref{thm:examples}(c), so $\rho_\epsilon$ is undistillable. It is NPT because $f(\rho_\epsilon)=-\epsilon a<0$.
\end{proof}

The proof gives $\epsilon_\infty$ explicitly. It works with any NPT bound entangled state whose partial transpose has a negative expectation value in $\ket{u_k}$, filtered with $k$ of the Schmidt coefficients. So when $\ket\varphi$ has Schmidt rank at least five, seven or eight, the states of Table~\ref{tab:wide} can give larger intervals, for example when $\ket\varphi$ is maximally entangled.

\subsection{A two-qutrit state with small integer entries}
\label{sec:279}

Local filters and separable noise turn two-qutrit Werner states into many other states. We give one with small integer entries. It is obtained from the Werner state $\tau_3=(31499\,I-10500\,\swap)/251991$, which has $p(\tau_3)=1/83997<p_0(3)$, so it is NPT and undistillable by Theorem~\ref{thm:family}. Let $P_i=\ket{i,i}\!\bra{i,i}$ on $\mathbb C^3\otimes\mathbb C^3$.

\begin{proposition}\label{prop:279}
The state
\[
 \rho=\frac{31\,I-14\,\swap+14\,P_2+28\,P_3}{279}
\]
is NPT and bound entangled. It is not a Werner state.
\end{proposition}

\begin{proof}
\emph{The state is NPT.} Let $R=279\,\rho$. On the span of $\ket{i,j}$ and $\ket{j,i}$ with $i\ne j$, the matrix $R$ is $\left(\begin{smallmatrix}31&-14\\-14&31\end{smallmatrix}\right)$, with the eigenvalues $17$ and $45$. On $\ket{1,1},\ket{2,2},\ket{3,3}$ it is diagonal with the entries $17,31,45$. So $R$ is positive definite, and its trace is $279$. By $\swap^\Gamma=\ket{u_3}\!\bra{u_3}$, the partial transpose $R^\Gamma$ acts on the span of $\ket{1,1},\ket{2,2},\ket{3,3}$ by
\[
 H=\begin{pmatrix}17&-14&-14\\-14&31&-14\\-14&-14&45\end{pmatrix}.
\]
It maps $(42,29,22)^T$ to $(0,3,-4)^T$, so $\ket v=42\ket{1,1}+29\ket{2,2}+22\ket{3,3}$ has $\bra vR^\Gamma\ket v=87-88=-1$. Since the diagonal entries of $\rho$ on $\ket{1,1}$ and $\ket{3,3}$ differ, $\rho$ does not commute with $U\otimes U$ for all unitaries $U$, and it is not a Werner state.

\emph{Undistillability.} Let $\kappa=3-1/m$ with $m=10500$, so that $\tau_3$ is proportional to $\kappa I-\swap$. Let $q=(7396,5095,3886)$, $\nu=4mq_3$, and let $A=\operatorname{diag}(6,7,8)$ on Alice's side and $B=\operatorname{diag}(B_1,B_2,B_3)$ with $B_i=1/(q_iA_i)$ on Bob's side. In Appendix~\ref{app:279} we show that
\begin{equation}\label{eq:279}
 R=\nu\,(A\otimes B)(\kappa I-\swap)(A\otimes B)+\sum_{i<j}\mu_{ij}\,S_{ij}+\Theta ,
\end{equation}
where the numbers $\mu_{ij}=14q_iq_j-\nu$ are positive, each $S_{ij}$ is an average of positive product operators, and $\Theta$ is diagonal in the product basis with nonnegative entries. The last two terms are separable. So Alice and Bob can prepare $\rho$ from $\tau_3$ by LOCC: they apply the local filters $A$ and $B$, rescaled to be Kraus operators of local measurements, and with suitable probabilities they replace the result by the normalized separable terms. By Theorem~\ref{thm:family}, $\tau_3$ is undistillable, so $\rho$ is undistillable. It is NPT, hence entangled, so it is bound entangled.
\end{proof}

\section{Discussion}

By Proposition~\ref{prop:key} and the remark after it, the Werner states of Theorem~\ref{thm:family} have zero distillable secret key under two-way LOCC. So do the states of Sec.~\ref{sec:nonwerner}, which we prepare from these Werner states by local filters and by mixing with separable states. The proof bounds the root fidelity between Eve's states in the two branches in which the bits of Alice and Bob are equal, while Eve keeps the public transcript. This is a consequence of the method of Sec.~\ref{sec:method} and of the argument of Ref.~\cite{openai}. Tabia and Hsieh~\cite{tabiahsieh}, Brand\~ao~\cite{brandao}, and Blakaj, Horodecki, Poszwa, Theurer and Winczewski~\cite[Theorem~1.1]{blakaj} noticed it as well.

We do not know the threshold $p_*(d)$ for any $d\ge3$. Our states lie close to the PPT boundary. In dimension three, the state $\rho_3$ of Theorem~\ref{thm:examples}(c) has $t=543/1579\approx0.343889$, and the boundary has $t=1/3$. In dimension eight, the state of Table~\ref{tab:intervals} has $t=4427/25000\approx0.177$, and the boundary has $t=1/8$. In both cases the conjecture reaches $t=1/2$. The method of Sec.~\ref{sec:method} gives only a sufficient condition for undistillability, and we do not know how far into the NPT region it can reach. In particular, we do not know whether the conjecture of Refs.~\cite{divincenzo,dur} holds, that is, whether the limit $c$ of Theorem~\ref{thm:threshold} is $1/2$. For two qutrits it says that the state $(2\,I-\swap)/15$, with $t=1/2$, is undistillable.

Our certificates grow as the dimension decreases, although the states stay simple. One reason is that the PPT boundary $t=1/d$ moves away from zero, so the maps must produce a larger coefficient of $\swap$ relative to $I$. With the input space $\mathbb C^d$, the maps of Sec.~\ref{sec:large} cannot do this in the dimensions three to five, because Lemma~\ref{lem:reverse} bounds the coefficient of $\swap$ from above (Sec.~\ref{sec:large}). The states of Sec.~\ref{sec:computer} use larger input spaces, which offer other covariant maps. The price is size. For an input space of dimension $r$, the Choi matrix has order $(rd)^2$. This order is $4096$ for the proof by hand in dimension eight, $313{,}600$ for the certificate in dimension eight, and $457{,}532{,}100$ for the two-qutrit certificate. In dimension eight, the input space $\wedge^4\mathbb C^8$ reaches $t=4427/25000$, while the input space $\mathbb C^8$ of the proof by hand reaches $t=6/47$. Symmetry reduces the check of $J\pm Q\succeq0$ for the two-qutrit certificate to $139$ blocks with at most $32$ rows, and the check of $Q^{\Gin}\succeq0$ to $145$ blocks. We do not know the smallest input space of a certificate for two qutrits.

Positivity under tensor powers was related to NPT bound entanglement before. M\"uller-Hermes, Reeb and Wolf call a linear map \emph{tensor-stable positive} if each of its tensor powers maps positive semidefinite matrices to positive semidefinite matrices. They proved that NPT bound entangled states exist if there is a tensor-stable positive map that is neither completely positive nor completely copositive~\cite[Theorem~4]{mrw}, and they state that the converse is open~\cite[Sec.~VI]{mrw}. We do not use their results, and our theorems do not decide whether such a map exists.

\appendix

\section{The matrix \texorpdfstring{$Q$}{Q} in dimension eight, by hand}
\label{app:hand}

We prove the rest of Lemma~\ref{lem:q8}: the matrix $J-Q$ is positive semidefinite, and $J+Q$ is positive definite. We use the notation of Secs.~\ref{sec:large} and~\ref{sec:eight}.

\step{Step 1: $J$ and $Q$ in terms of $F_{13}$, $F_{14}$ and the maps $T_k$} The definitions of the maps $T_k$ give
\begin{equation}\label{eq:moves}
\begin{aligned}
 F_{13}T_0&=T_1, & F_{13}T_1&=T_0, & F_{13}T_2&=T_2F,\\
 F_{14}T_0&=T_2, & F_{14}T_2&=T_0, & F_{14}T_1&=T_1F,
\end{aligned}
\end{equation}
and the products \eqref{eq:gram-large} with $d=8$:
\begin{equation}\label{eq:gram}
 T_k^\dagger T_k=8\,I,\qquad T_0^\dagger T_1=T_0^\dagger T_2=I,\qquad T_1^\dagger T_2=F .
\end{equation}
The Choi matrix of $L$ is $5\,I-3\,F_{13}$. The Choi matrix of $M$ is $I+2\,\ket u\!\bra u$ on the systems $2,4$, and $\ket u\!\bra u$ on these systems is $T_2T_2^\dagger$. With $F_{13}T_2=T_2F$,
\begin{equation}\label{eq:Jhand}
\begin{aligned}
 J&=(5\,I-3\,F_{13})(I+2\,T_2T_2^\dagger)\\
 &=5\,I-3\,F_{13}+10\,T_2T_2^\dagger-6\,T_2FT_2^\dagger .
\end{aligned}
\end{equation}

Next we write $Q^{\Gin}$ as a sum of squares. Let $\Pi_\pm=(I\pm F)/2$. Then $R=47\,I-6\,F=41\,\Pi_++53\,\Pi_-$, so $R^{-1}=\Pi_+/41+\Pi_-/53$. For the input basis vector $\ket{i,j}$, Step~5 of Sec.~\ref{sec:large}, with $s=3/5$, $b=2$ and the factor $5$ of $L$, gives the block
\[
 W_{ij}=5\,\delta_{ij}\,I-3\,E^{(3)}_{ji}+10\,E^{(4)}_{ji}-6\,E^{(4)}_{ji}F
\]
of $JV$. So $W_{ij}^\dagger=5\,\delta_{ij}\,I-3\,E^{(3)}_{ij}+10\,E^{(4)}_{ij}-6\,FE^{(4)}_{ij}$. We write out the four sums over $i,j$. For example, $\sum_{i,j}\ket{i,j}\otimes E^{(3)}_{ij}$ maps $\ket{x,y}$ to $\sum_i\ket{i,x,i,y}=ST_1\ket{x,y}$. In the same way,
\[
 \sum_{i,j=1}^{8}\ket{i,j}\otimes W_{ij}^\dagger=S\,B,\qquad B=5\,T_0-3\,T_1+10\,T_2-6\,T_1F .
\]
In the proof of Lemma~\ref{lem:form} we saw that $N^{\Gin}=\mathcal WR^{-1}\mathcal W^\dagger$ with $\mathcal W=\sum_p\ket p\otimes W_p^\dagger$. Here $\mathcal W=SB$, so $N^{\Gin}=SBR^{-1}B^\dagger S$. Since $F\Pi_\pm=\pm\Pi_\pm$, we have $B\Pi_\pm=A_\pm\Pi_\pm$ with $A_+=5T_0-9T_1+10T_2$ and $A_-=5T_0+3T_1+10T_2$. Also $ET_2=T_2-T_0/8$ by \eqref{eq:gram}, so $H=A_0\Pi_-A_0^\dagger$ with $A_0=T_2-T_0/8$. By \eqref{eq:Q8},
\begin{equation}\label{eq:squares}
 S\,Q^{\Gin}S=\frac{A_+\Pi_+A_+^\dagger}{41}+\frac{A_-\Pi_-A_-^\dagger}{53}+A_0\Pi_-A_0^\dagger .
\end{equation}
Each term is a sum of terms $T_m\Pi_\pm T_n^\dagger$. With $\Pi_\pm=(I\pm F)/2$, the right-hand side is $\sum_{m,n=0}^{2}T_m(X_{mn}I+Y_{mn}F)T_n^\dagger$, where $X=(k_++k_-)/2$, $Y=(k_+-k_-)/2$,
\[
 k_+=\frac{a_+a_+^{T}}{41},\qquad k_-=\frac{a_-a_-^{T}}{53}+a_0a_0^{T},
\]
and $a_+=(5,-9,10)^T$, $a_-=(5,3,10)^T$ and $a_0=(-1/8,0,1)^T$ are the coefficients of $A_+$, $A_-$ and $A_0$.

We get $Q$ back by applying the map $X\mapsto S\,X^{\Gin}S$ to \eqref{eq:squares}, because the input transpose commutes with $S$. This map permutes the operators that occur. Writing out the entries of each operator shows that it exchanges
\begin{gather*}
 T_1T_1^\dagger\leftrightarrow F_{13},\qquad T_2T_2^\dagger\leftrightarrow F_{14},\\
 T_2FT_1^\dagger\leftrightarrow F_{13}F_{14},\qquad T_1FT_2^\dagger\leftrightarrow F_{14}F_{13},\\
 T_1FT_1^\dagger\leftrightarrow T_2FT_2^\dagger,\qquad T_0FT_1^\dagger\leftrightarrow T_2FT_0^\dagger,\\
 T_0FT_2^\dagger\leftrightarrow T_1FT_0^\dagger,
\end{gather*}
that it maps $T_mT_n^\dagger$ to $T_nT_m^\dagger$ for $m\ne n$, and that it fixes $T_0T_0^\dagger$ and $T_0FT_0^\dagger$. For example, $T_1T_1^\dagger$ has the entry $\delta_{i_1j_1}\delta_{i_2i_3}\delta_{j_2j_3}\delta_{i_4j_4}$ in the row $(i_1,i_2,i_3,i_4)$ and the column $(j_1,j_2,j_3,j_4)$. The input transpose exchanges $i_1,i_2$ with $j_1,j_2$, and $S$ then exchanges the positions $1$ and $2$ in the row and in the column. This gives $\delta_{i_1j_3}\delta_{i_2j_2}\delta_{i_3j_1}\delta_{i_4j_4}$, the entry of $F_{13}$. Since $X$ and $Y$ are symmetric, we get
\begin{equation}\label{eq:Qhand}
\begin{aligned}
 Q={}&\alpha\,F_{13}+\beta\,(F_{13}F_{14}+F_{14}F_{13})+\gamma\,F_{14}\\
 &+\sum_{m,n=0}^{2}T_m\,(c_{mn}I+c'_{mn}F)\,T_n^\dagger ,
\end{aligned}
\end{equation}
with $\alpha=X_{11}$, $\beta=Y_{12}$, $\gamma=X_{22}$ and the symmetric matrices
\[
 c=\begin{pmatrix}X_{00}&X_{01}&X_{02}\\X_{01}&0&X_{12}\\X_{02}&X_{12}&0\end{pmatrix},\qquad
 c'=\begin{pmatrix}Y_{00}&Y_{02}&Y_{01}\\Y_{02}&Y_{22}&0\\Y_{01}&0&Y_{11}\end{pmatrix}.
\]
In numbers, with $2173=41\cdot53$, we have $\alpha=2331/2173$, $\beta=-3000/2173$, $\gamma=11573/4346$, and
\begin{align*}
 c_{00}&=\tfrac{152573}{278144}, & c_{01}&=-\tfrac{885}{2173}, & c_{02}&=\tfrac{35427}{34768},\\
 c_{12}&=-\tfrac{1770}{2173}, & c'_{00}&=\tfrac{17027}{278144}, & c'_{01}&=\tfrac{6973}{34768},\\
 c'_{02}&=-\tfrac{1500}{2173}, & c'_{11}&=-\tfrac{973}{4346}, & c'_{22}&=\tfrac{1962}{2173},
\end{align*}
while $c_{11}=c_{22}=c'_{12}=0$.

\step{Step 2: two invariant subspaces} Let $\mathcal H_1$ be the span of the ranges of $T_0$, $T_1$ and $T_2$. By \eqref{eq:moves}, $F_{13}$ and $F_{14}$ map $\mathcal H_1$ into itself. The terms with a map $T_k$ send every vector into $\mathcal H_1$. So $J$ and $Q$ map $\mathcal H_1$ into itself. They are Hermitian, so they also map the orthogonal complement $\mathcal H_1^\perp$ into itself. We check the two inequalities on $\mathcal H_1^\perp$ and on $\mathcal H_1$.

\step{Step 3: the complement} A vector of $\mathcal H_1^\perp$ is annihilated by every $T_n^\dagger$. So on $\mathcal H_1^\perp$ only the terms without maps $T_k$ remain: $J=5\,I-3\,F_{13}$ and $Q=\alpha F_{13}+\beta\,(F_{13}F_{14}+F_{14}F_{13})+\gamma F_{14}$. These are combinations of the six permutations of the systems $1,3,4$. Every unitary representation of this group of permutations is a sum of copies of three irreducible ones, and we check the inequalities in each of them. We use the bounds
\[
 1<\alpha<\tfrac{13}{12},\qquad -\tfrac75<\beta<-\tfrac43,\qquad \tfrac52<\gamma<\tfrac83 .
\]

In the trivial representation every permutation is $1$. Then $J=2$ and $Q=\alpha+2\beta+\gamma=4235/4346$, so $|Q|<J$.

In the sign representation every exchange is $-1$, and the two products $F_{13}F_{14}$ and $F_{14}F_{13}$ are $1$. Then $J=8$ and $Q=-\alpha+2\beta-\gamma=-28235/4346$, so $|Q|<J$.

In the representation of dimension two, the three exchanges are reflections of the plane, and any two of them, $F$ and $F'$, satisfy $FF'+F'F=-I$. Their traces are zero. So for real numbers $x,y$ the matrix $X=x\,F_{13}+y\,F_{14}$ has trace zero and $X^2=(x^2+y^2-xy)\,I$. Its eigenvalues are $\pm\sqrt{x^2+y^2-xy}$, and $z\,I+X$ is positive definite exactly when $z>0$ and $z^2>x^2+y^2-xy$. In this representation $F_{13}F_{14}+F_{14}F_{13}=-I$, so
\begin{align*}
 J-Q&=(5+\beta)\,I-(3+\alpha)\,F_{13}-\gamma\,F_{14},\\
 J+Q&=(5-\beta)\,I-(3-\alpha)\,F_{13}+\gamma\,F_{14}.
\end{align*}
For $J-Q$ we have $z=5+\beta>18/5$, $x=-(3+\alpha)$ and $y=-\gamma$. The function $x^2+y^2-xy$ increases with $|x|$ and with $|y|$ when $2|x|>|y|$ and $2|y|>|x|$, which holds for $4<|x|<49/12$ and $5/2<|y|<8/3$. So
\[
 x^2+y^2-xy<\big(\tfrac{49}{12}\big)^2+\big(\tfrac83\big)^2-\tfrac{49}{12}\cdot\tfrac83=\tfrac{619}{48}<\tfrac{324}{25}=\big(\tfrac{18}{5}\big)^2 .
\]
For $J+Q$ we have $z=5-\beta>19/3$, $x=-(3-\alpha)$ and $y=\gamma$, and $x^2+y^2-xy=(3-\alpha)^2+\gamma^2+(3-\alpha)\gamma<4+\tfrac{64}9+\tfrac{16}3=\tfrac{148}9<\big(\tfrac{19}3\big)^2$. So $J-Q$ and $J+Q$ are positive definite on $\mathcal H_1^\perp$.

\step{Step 4: the subspace $\mathcal H_1$} Every vector of $\mathcal H_1$ is $T_0x_0+T_1x_1+T_2x_2$ with vectors $x_k$ of the output. By \eqref{eq:gram}, the products $T_k^\dagger T_l$ are numbers on the symmetric vectors, where $F=1$, and on the antisymmetric vectors, where $F=-1$. For $\varepsilon=\pm1$ they form the matrix
\[
 G_\varepsilon=\begin{pmatrix}8&1&1\\1&8&\varepsilon\\1&\varepsilon&8\end{pmatrix},
\]
which is positive definite. By \eqref{eq:moves}, $F_{13}$ and $F_{14}$ act on the three maps by the matrices
\[
 a_\varepsilon=\begin{pmatrix}0&1&0\\1&0&0\\0&0&\varepsilon\end{pmatrix},\qquad
 b_\varepsilon=\begin{pmatrix}0&0&1\\0&\varepsilon&0\\1&0&0\end{pmatrix}.
\]
A term $T_m\,(c\,I+c'F)\,T_n^\dagger$ acts by the matrix that has the row $n$ of $G_\varepsilon$, multiplied by $c+\varepsilon c'$, in its row $m$, and zeros elsewhere. So the sum in \eqref{eq:Qhand} acts by $(c+\varepsilon c')\,G_\varepsilon$. An operator $O$ that maps $\mathcal H_1$ into itself and acts on the three maps by the matrix $O_\varepsilon$ has the quadratic form $G_\varepsilon O_\varepsilon$ on vectors of symmetry $\varepsilon$. So $O$ is positive semidefinite on $\mathcal H_1$ exactly when the two $3\times3$ matrices $G_+O_+$ and $G_-O_-$ are positive semidefinite. For $J$, \eqref{eq:Jhand} gives
\begin{align*}
 G_+J_+&=\begin{pmatrix}41&-15&34\\-15&41&34\\34&34&272\end{pmatrix},\\
 G_-J_-&=\begin{pmatrix}53&-35&136\\-35&53&-136\\136&-136&1088\end{pmatrix}.
\end{align*}

We use one test for symmetric $2\times2$ matrices. If $2x>|y|$ and $w>2|y|$, then $xw>y^2$, and the matrix with the rows $(x,y)$ and $(y,w)$ is positive definite.

\emph{The inequality $J-Q\succeq0$.} Since $(J-Q)T_0=0$, the first column of $(J_\varepsilon-Q_\varepsilon)$ is zero, and so are the first row and the first column of $G_\varepsilon(J_\varepsilon-Q_\varepsilon)$. We compute $G_\varepsilon Q_\varepsilon$ from \eqref{eq:Qhand}. The remaining $2\times2$ blocks of $G_\varepsilon(J_\varepsilon-Q_\varepsilon)$ are
\begin{align*}
 \varepsilon=+1:&\quad \tfrac{7}{17384}\begin{pmatrix}123883&219772\\219772&530293\end{pmatrix},\\
 \varepsilon=-1:&\quad \tfrac{9}{139072}\begin{pmatrix}875651&-1402397\\-1402397&17100227\end{pmatrix}.
\end{align*}
Both pass the test: $2\cdot123883>219772$ and $530293>2\cdot219772$, and $2\cdot875651>1402397$ and $17100227>2\cdot1402397$.

\emph{The inequality $J+Q\succ0$.} We use $J+Q=2J-(J-Q)$. So $G_\varepsilon(J_\varepsilon+Q_\varepsilon)$ has the first row $(82,-30,68)$ for $\varepsilon=+1$ and $(106,-70,272)$ for $\varepsilon=-1$, and its lower $2\times2$ block is twice the block of $G_\varepsilon J_\varepsilon$ minus the matrix above. A symmetric matrix with a positive first entry is positive definite exactly when the Schur complement of this entry is positive definite. The Schur complements are
\begin{align*}
 \varepsilon=+1:&\quad \tfrac{7}{17384}\begin{pmatrix}52501&10884\\10884&680651\end{pmatrix},\\
 \varepsilon=-1:&\quad \tfrac{27}{139072}\begin{pmatrix}15999&-8353\\-8353&1913023\end{pmatrix},
\end{align*}
and both pass the test. Together with Step~3, this proves Lemma~\ref{lem:q8}.

\section{The decomposition of the two-qutrit state}
\label{app:279}

We prove Eq.~\eqref{eq:279}, which is used in the proof of Proposition~\ref{prop:279}. We use the notation of that proof. Let $H$ be the $3\times3$ matrix of that proof.

\step{The filter} The vector $q$ satisfies $Hq=-(2,3,4)^T$, and $A_iB_i=1/q_i$. Let
\[
 R_0=\nu\,(A\otimes B)(\kappa I-\swap)(A\otimes B).
\]
Its entries are $\nu(\kappa-\delta_{ij})A_i^2B_j^2$ on the diagonal at $\ket{i,j}$, and $-\nu/(q_iq_j)$ between $\ket{i,j}$ and $\ket{j,i}$ for $i\ne j$. All other entries are zero.

\step{Separable terms} For $i<j$ let $S_{ij}$ be the average of $U\ket x\!\bra xU^\dagger\otimes U\ket y\!\bra yU^\dagger$, where $\ket x=A_i\ket i+A_j\ket j$, $\ket y=B_i\ket i-B_j\ket j$, and $U$ is diagonal, with independent random fourth roots of unity as its diagonal entries. It is separable. The average keeps the entry between $\ket{a,b}$ and $\ket{c,e}$ only when $\{a,b\}=\{c,e\}$. So $S_{ij}$ has the entry $-1/(q_iq_j)$ between $\ket{i,j}$ and $\ket{j,i}$, the diagonal entries $1/q_i^2$, $1/q_j^2$ at $\ket{i,i}$, $\ket{j,j}$ and $A_i^2B_j^2$, $A_j^2B_i^2$ at $\ket{i,j}$, $\ket{j,i}$, and no other entries. Let $w_{ij}=14q_iq_j$, so that $\mu_{ij}=w_{ij}-\nu$, and let
\[
 R_1=R_0+\sum_{i<j}\mu_{ij}\,S_{ij}.
\]
The weights are positive, because $w_{ij}\ge14q_2q_3$ and $14q_2>4m$. The entry of $R_1$ between $\ket{i,j}$ and $\ket{j,i}$ is $-(\nu+\mu_{ij})/(q_iq_j)=-14$, as in $R$.

\step{The diagonal} We show that $R-R_1$ is diagonal with nonnegative entries. Its entries off the diagonal are zero by the last step. At $\ket{i,i}$, the matrix $R_0$ contributes $\nu(\kappa-1)/q_i^2$, and the two terms $S_{ij}$ with $j\ne i$ contribute $\sum_{j\ne i}(w_{ij}-\nu)/q_i^2$. Since $\nu(\kappa-1)-2\nu=-\nu/m$, the entry of $R_1$ is $(\sum_{j\ne i}w_{ij}-\nu/m)/q_i^2$. Since $H_{ij}=-14$ for $i\ne j$, this is at most $H_{ii}$ exactly when $\nu/m\ge-q_i(Hq)_i$. The right-hand sides are $14792$, $15285$ and $15544$, and $\nu/m=4q_3=15544$. At $\ket{i,j}$ with $i\ne j$, let $g_i=q_iA_i^2$, so that $g=(266256,249655,248704)$. The entry of $R_1$ is
\[
 \Big(14+\frac{(\kappa-1)\nu}{q_iq_j}\Big)\frac{g_i}{g_j}.
\]
We have $(\kappa-1)\nu<8mq_3=84000\,q_3$. For the pairs $\{1,2\}$ and $\{1,3\}$ the first factor is less than $26$, and both ratios $g_i/g_j$ and $g_j/g_i$ are less than $15/14$. For the pair $\{2,3\}$ the first factor is less than $61/2$, and both ratios are less than $101/100$. So each entry is less than $31$, the entry of $R$. Hence $\Theta=R-R_1$ is diagonal with nonnegative entries, which is \eqref{eq:279}.

\bibliography{paper}

\begin{thebibliography}{39}%
\makeatletter
\providecommand \@ifxundefined [1]{%
 \@ifx{#1\undefined}
}%
\providecommand \@ifnum [1]{%
 \ifnum #1\expandafter \@firstoftwo
 \else \expandafter \@secondoftwo
 \fi
}%
\providecommand \@ifx [1]{%
 \ifx #1\expandafter \@firstoftwo
 \else \expandafter \@secondoftwo
 \fi
}%
\providecommand \natexlab [1]{#1}%
\providecommand \enquote  [1]{``#1''}%
\providecommand \bibnamefont  [1]{#1}%
\providecommand \bibfnamefont [1]{#1}%
\providecommand \citenamefont [1]{#1}%
\providecommand \href@noop [0]{\@secondoftwo}%
\providecommand \href [0]{\begingroup \@sanitize@url \@href}%
\providecommand \@href[1]{\@@startlink{#1}\@@href}%
\providecommand \@@href[1]{\endgroup#1\@@endlink}%
\providecommand \@sanitize@url [0]{\catcode `\\12\catcode `\$12\catcode
  `\&12\catcode `\#12\catcode `\^12\catcode `\_12\catcode `\%12\relax}%
\providecommand \@@startlink[1]{}%
\providecommand \@@endlink[0]{}%
\providecommand \url  [0]{\begingroup\@sanitize@url \@url }%
\providecommand \@url [1]{\endgroup\@href {#1}{\urlprefix }}%
\providecommand \urlprefix  [0]{URL }%
\providecommand \Eprint [0]{\href }%
\providecommand \doibase [0]{https://doi.org/}%
\providecommand \selectlanguage [0]{\@gobble}%
\providecommand \bibinfo  [0]{\@secondoftwo}%
\providecommand \bibfield  [0]{\@secondoftwo}%
\providecommand \translation [1]{[#1]}%
\providecommand \BibitemOpen [0]{}%
\providecommand \bibitemStop [0]{}%
\providecommand \bibitemNoStop [0]{.\EOS\space}%
\providecommand \EOS [0]{\spacefactor3000\relax}%
\providecommand \BibitemShut  [1]{\csname bibitem#1\endcsname}%
\let\auto@bib@innerbib\@empty
\bibitem [{\citenamefont {Bandyopadhyay}(2026)}]{bandyopadhyay-note}%
  \BibitemOpen
  \bibfield  {author} {\bibinfo {author} {\bibfnamefont {S.}~\bibnamefont
  {Bandyopadhyay}},\ }\href@noop {} {\bibinfo {title} {A remark on the {W}erner
  state of {arXiv}:2611.34921}},\ \bibinfo {howpublished} {Personal
  communication, received by email} (\bibinfo {year} {2026}),\ \bibinfo {note}
  {{November} 26, 2026}\BibitemShut {NoStop}%
\bibitem [{\citenamefont {Bandyopadhyay}\ and\ \citenamefont
  {Roychowdhury}(2003)}]{br03}%
  \BibitemOpen
  \bibfield  {author} {\bibinfo {author} {\bibfnamefont {S.}~\bibnamefont
  {Bandyopadhyay}}\ and\ \bibinfo {author} {\bibfnamefont {V.}~\bibnamefont
  {Roychowdhury}},\ }\bibfield  {title} {\bibinfo {title} {New classes of
  $n$-copy undistillable quantum states with negative partial transposition},\
  }\href {https://doi.org/10.1103/PhysRevA.68.022319} {\bibfield  {journal}
  {\bibinfo  {journal} {Phys. Rev. A}\ }\textbf {\bibinfo {volume} {68}},\
  \bibinfo {pages} {022319} (\bibinfo {year} {2003})},\ \bibinfo {note}
  {arXiv:quant-ph/0302093}\BibitemShut {NoStop}%
\bibitem [{\citenamefont {Tabia}\ and\ \citenamefont
  {Hsieh}(2026)}]{tabiahsieh}%
  \BibitemOpen
  \bibfield  {author} {\bibinfo {author} {\bibfnamefont {G.~N.~M.}\
  \bibnamefont {Tabia}}\ and\ \bibinfo {author} {\bibfnamefont {M.-H.}\
  \bibnamefont {Hsieh}},\ }\href@noop {} {}\bibinfo {howpublished} {Personal
  communication} (\bibinfo {year} {2026}),\ \bibinfo {note} {{November} 2026,
  will be made available at
  \url{https://github.com/gelonoel/npt-zero-key}}\BibitemShut {NoStop}%
\bibitem [{\citenamefont {Brand{\~a}o}(2026)}]{brandao}%
  \BibitemOpen
  \bibfield  {author} {\bibinfo {author} {\bibfnamefont {F.}~\bibnamefont
  {Brand{\~a}o}},\ }\href@noop {} {}\bibinfo {howpublished} {Personal
  communication} (\bibinfo {year} {2026}),\ \bibinfo {note} {{November}
  2026}\BibitemShut {NoStop}%
\bibitem [{\citenamefont {{OpenAI}}(2026)}]{openai}%
  \BibitemOpen
  \bibfield  {author} {\bibinfo {author} {\bibnamefont {{OpenAI}}},\ }\href
  {https://github.com/openai/math/blob/main/preprints/Entanglement-with-zero-distillable-secret-key-in-local-dimension-ten-September-27-2026/paper.pdf}
  {\bibinfo {title} {Entanglement with zero distillable secret key in local
  dimension ten}} (\bibinfo {year} {2026}),\ \bibinfo {note} {preprint, 27
  September 2026}\BibitemShut {NoStop}%
\bibitem [{\citenamefont {Blakaj}\ \emph {et~al.}(2026)\citenamefont {Blakaj},
  \citenamefont {Horodecki}, \citenamefont {Poszwa}, \citenamefont {Theurer},\
  and\ \citenamefont {Winczewski}}]{blakaj}%
  \BibitemOpen
  \bibfield  {author} {\bibinfo {author} {\bibfnamefont {V.}~\bibnamefont
  {Blakaj}}, \bibinfo {author} {\bibfnamefont {K.}~\bibnamefont {Horodecki}},
  \bibinfo {author} {\bibfnamefont {S.}~\bibnamefont {Poszwa}}, \bibinfo
  {author} {\bibfnamefont {T.}~\bibnamefont {Theurer}},\ and\ \bibinfo {author}
  {\bibfnamefont {M.}~\bibnamefont {Winczewski}},\ }\href@noop {} {\bibinfo
  {title} {{NPT} states with zero distillable secret key}} (\bibinfo {year}
  {2026}),\ \Eprint {https://arxiv.org/abs/2611.37775} {arXiv:2611.37775}
  \BibitemShut {NoStop}%
\bibitem [{\citenamefont {Wu}(2026)}]{wu}%
  \BibitemOpen
  \bibfield  {author} {\bibinfo {author} {\bibfnamefont {Y.-C.}\ \bibnamefont
  {Wu}},\ }\href@noop {} {}\bibinfo {howpublished} {Personal communication}
  (\bibinfo {year} {2026})\BibitemShut {NoStop}%
\bibitem [{\citenamefont {Eisert}(2026)}]{eisert}%
  \BibitemOpen
  \bibfield  {author} {\bibinfo {author} {\bibfnamefont {J.}~\bibnamefont
  {Eisert}},\ }\href@noop {} {}\bibinfo {howpublished} {Personal communication}
  (\bibinfo {year} {2026})\BibitemShut {NoStop}%
\bibitem [{\citenamefont {Peres}(1996)}]{peres}%
  \BibitemOpen
  \bibfield  {author} {\bibinfo {author} {\bibfnamefont {A.}~\bibnamefont
  {Peres}},\ }\bibfield  {title} {\bibinfo {title} {Separability criterion for
  density matrices},\ }\href {https://doi.org/10.1103/PhysRevLett.77.1413}
  {\bibfield  {journal} {\bibinfo  {journal} {Phys. Rev. Lett.}\ }\textbf
  {\bibinfo {volume} {77}},\ \bibinfo {pages} {1413} (\bibinfo {year}
  {1996})}\BibitemShut {NoStop}%
\bibitem [{\citenamefont {Horodecki}\ \emph {et~al.}(1998)\citenamefont
  {Horodecki}, \citenamefont {Horodecki},\ and\ \citenamefont
  {Horodecki}}]{horodecki98}%
  \BibitemOpen
  \bibfield  {author} {\bibinfo {author} {\bibfnamefont {M.}~\bibnamefont
  {Horodecki}}, \bibinfo {author} {\bibfnamefont {P.}~\bibnamefont
  {Horodecki}},\ and\ \bibinfo {author} {\bibfnamefont {R.}~\bibnamefont
  {Horodecki}},\ }\bibfield  {title} {\bibinfo {title} {Mixed-state
  entanglement and distillation: Is there a ``bound'' entanglement in
  nature?},\ }\href {https://doi.org/10.1103/PhysRevLett.80.5239} {\bibfield
  {journal} {\bibinfo  {journal} {Phys. Rev. Lett.}\ }\textbf {\bibinfo
  {volume} {80}},\ \bibinfo {pages} {5239} (\bibinfo {year}
  {1998})}\BibitemShut {NoStop}%
\bibitem [{\citenamefont {Smolin}(2001)}]{smolin}%
  \BibitemOpen
  \bibfield  {author} {\bibinfo {author} {\bibfnamefont {J.~A.}\ \bibnamefont
  {Smolin}},\ }\bibfield  {title} {\bibinfo {title} {Four-party unlockable
  bound entangled state},\ }\href {https://doi.org/10.1103/PhysRevA.63.032306}
  {\bibfield  {journal} {\bibinfo  {journal} {Phys. Rev. A}\ }\textbf {\bibinfo
  {volume} {63}},\ \bibinfo {pages} {032306} (\bibinfo {year}
  {2001})}\BibitemShut {NoStop}%
\bibitem [{\citenamefont {DiVincenzo}\ \emph {et~al.}(2000)\citenamefont
  {DiVincenzo}, \citenamefont {Shor}, \citenamefont {Smolin}, \citenamefont
  {Terhal},\ and\ \citenamefont {Thapliyal}}]{divincenzo}%
  \BibitemOpen
  \bibfield  {author} {\bibinfo {author} {\bibfnamefont {D.~P.}\ \bibnamefont
  {DiVincenzo}}, \bibinfo {author} {\bibfnamefont {P.~W.}\ \bibnamefont
  {Shor}}, \bibinfo {author} {\bibfnamefont {J.~A.}\ \bibnamefont {Smolin}},
  \bibinfo {author} {\bibfnamefont {B.~M.}\ \bibnamefont {Terhal}},\ and\
  \bibinfo {author} {\bibfnamefont {A.~V.}\ \bibnamefont {Thapliyal}},\
  }\bibfield  {title} {\bibinfo {title} {Evidence for bound entangled states
  with negative partial transpose},\ }\href
  {https://doi.org/10.1103/PhysRevA.61.062312} {\bibfield  {journal} {\bibinfo
  {journal} {Phys. Rev. A}\ }\textbf {\bibinfo {volume} {61}},\ \bibinfo
  {pages} {062312} (\bibinfo {year} {2000})}\BibitemShut {NoStop}%
\bibitem [{\citenamefont {D{\"u}r}\ \emph {et~al.}(2000)\citenamefont
  {D{\"u}r}, \citenamefont {Cirac}, \citenamefont {Lewenstein},\ and\
  \citenamefont {Bru{\ss}}}]{dur}%
  \BibitemOpen
  \bibfield  {author} {\bibinfo {author} {\bibfnamefont {W.}~\bibnamefont
  {D{\"u}r}}, \bibinfo {author} {\bibfnamefont {J.~I.}\ \bibnamefont {Cirac}},
  \bibinfo {author} {\bibfnamefont {M.}~\bibnamefont {Lewenstein}},\ and\
  \bibinfo {author} {\bibfnamefont {D.}~\bibnamefont {Bru{\ss}}},\ }\bibfield
  {title} {\bibinfo {title} {Distillability and partial transposition in
  bipartite systems},\ }\href {https://doi.org/10.1103/PhysRevA.61.062313}
  {\bibfield  {journal} {\bibinfo  {journal} {Phys. Rev. A}\ }\textbf {\bibinfo
  {volume} {61}},\ \bibinfo {pages} {062313} (\bibinfo {year}
  {2000})}\BibitemShut {NoStop}%
\bibitem [{\citenamefont {Werner}(1989)}]{werner}%
  \BibitemOpen
  \bibfield  {author} {\bibinfo {author} {\bibfnamefont {R.~F.}\ \bibnamefont
  {Werner}},\ }\bibfield  {title} {\bibinfo {title} {Quantum states with
  {Einstein-Podolsky-Rosen} correlations admitting a hidden-variable model},\
  }\href {https://doi.org/10.1103/PhysRevA.40.4277} {\bibfield  {journal}
  {\bibinfo  {journal} {Phys. Rev. A}\ }\textbf {\bibinfo {volume} {40}},\
  \bibinfo {pages} {4277} (\bibinfo {year} {1989})}\BibitemShut {NoStop}%
\bibitem [{\citenamefont {Simon}(2006)}]{simon}%
  \BibitemOpen
  \bibfield  {author} {\bibinfo {author} {\bibfnamefont {R.}~\bibnamefont
  {Simon}},\ }\href@noop {} {\bibinfo {title} {{NPPT} bound entanglement
  exists}} (\bibinfo {year} {2006}),\ \Eprint
  {https://arxiv.org/abs/quant-ph/0608250} {arXiv:quant-ph/0608250}
  \BibitemShut {NoStop}%
\bibitem [{\citenamefont {Chattopadhyay}\ and\ \citenamefont
  {Sarkar}(2006)}]{chattopadhyay}%
  \BibitemOpen
  \bibfield  {author} {\bibinfo {author} {\bibfnamefont {I.}~\bibnamefont
  {Chattopadhyay}}\ and\ \bibinfo {author} {\bibfnamefont {D.}~\bibnamefont
  {Sarkar}},\ }\href@noop {} {\bibinfo {title} {{NPT} bound entanglement---the
  problem revisited}} (\bibinfo {year} {2006}),\ \Eprint
  {https://arxiv.org/abs/quant-ph/0609050} {arXiv:quant-ph/0609050}
  \BibitemShut {NoStop}%
\bibitem [{\citenamefont {Hiroshima}(2008)}]{hiroshima}%
  \BibitemOpen
  \bibfield  {author} {\bibinfo {author} {\bibfnamefont {T.}~\bibnamefont
  {Hiroshima}},\ }\href@noop {} {\bibinfo {title} {Bound entangled states with
  non-positive partial transpose exist}} (\bibinfo {year} {2008}),\ \bibinfo
  {note} {withdrawn by the author},\ \Eprint {https://arxiv.org/abs/0802.0329}
  {arXiv:0802.0329} \BibitemShut {NoStop}%
\bibitem [{\citenamefont {Pankowski}\ \emph {et~al.}(2010)\citenamefont
  {Pankowski}, \citenamefont {Piani}, \citenamefont {Horodecki},\ and\
  \citenamefont {Horodecki}}]{pankowski}%
  \BibitemOpen
  \bibfield  {author} {\bibinfo {author} {\bibfnamefont {{\L}.}~\bibnamefont
  {Pankowski}}, \bibinfo {author} {\bibfnamefont {M.}~\bibnamefont {Piani}},
  \bibinfo {author} {\bibfnamefont {M.}~\bibnamefont {Horodecki}},\ and\
  \bibinfo {author} {\bibfnamefont {P.}~\bibnamefont {Horodecki}},\ }\bibfield
  {title} {\bibinfo {title} {A few steps more towards {NPT} bound
  entanglement},\ }\href {https://doi.org/10.1109/TIT.2010.2050810} {\bibfield
  {journal} {\bibinfo  {journal} {IEEE Trans. Inf. Theory}\ }\textbf {\bibinfo
  {volume} {56}},\ \bibinfo {pages} {4085} (\bibinfo {year}
  {2010})}\BibitemShut {NoStop}%
\bibitem [{\citenamefont {Horodecki}\ \emph {et~al.}(2022)\citenamefont
  {Horodecki}, \citenamefont {Rudnicki},\ and\ \citenamefont
  {{\.Z}yczkowski}}]{fiveopen}%
  \BibitemOpen
  \bibfield  {author} {\bibinfo {author} {\bibfnamefont {P.}~\bibnamefont
  {Horodecki}}, \bibinfo {author} {\bibfnamefont {{\L}.}~\bibnamefont
  {Rudnicki}},\ and\ \bibinfo {author} {\bibfnamefont {K.}~\bibnamefont
  {{\.Z}yczkowski}},\ }\bibfield  {title} {\bibinfo {title} {Five open problems
  in quantum information theory},\ }\href
  {https://doi.org/10.1103/PRXQuantum.3.010101} {\bibfield  {journal} {\bibinfo
   {journal} {PRX Quantum}\ }\textbf {\bibinfo {volume} {3}},\ \bibinfo {pages}
  {010101} (\bibinfo {year} {2022})}\BibitemShut {NoStop}%
\bibitem [{\citenamefont {Qi}\ \emph {et~al.}(2024)\citenamefont {Qi},
  \citenamefont {Gupur}, \citenamefont {Wu},\ and\ \citenamefont {Guo}}]{qi}%
  \BibitemOpen
  \bibfield  {author} {\bibinfo {author} {\bibfnamefont {S.-Y.}\ \bibnamefont
  {Qi}}, \bibinfo {author} {\bibfnamefont {G.}~\bibnamefont {Gupur}}, \bibinfo
  {author} {\bibfnamefont {Y.-C.}\ \bibnamefont {Wu}},\ and\ \bibinfo {author}
  {\bibfnamefont {G.-P.}\ \bibnamefont {Guo}},\ }\bibfield  {title} {\bibinfo
  {title} {Unveiling {NPT} bound problem: From distillability sets to
  inequalities and multivariable insights},\ }\href
  {https://doi.org/10.1103/PhysRevA.110.012406} {\bibfield  {journal} {\bibinfo
   {journal} {Phys. Rev. A}\ }\textbf {\bibinfo {volume} {110}},\ \bibinfo
  {pages} {012406} (\bibinfo {year} {2024})},\ \bibinfo {note}
  {arXiv:2402.18037}\BibitemShut {NoStop}%
\bibitem [{\citenamefont {Fu}\ \emph {et~al.}(2026)\citenamefont {Fu},
  \citenamefont {Gao},\ and\ \citenamefont {Park}}]{fur}%
  \BibitemOpen
  \bibfield  {author} {\bibinfo {author} {\bibfnamefont {J.}~\bibnamefont
  {Fu}}, \bibinfo {author} {\bibfnamefont {L.}~\bibnamefont {Gao}},\ and\
  \bibinfo {author} {\bibfnamefont {S.-J.}\ \bibnamefont {Park}},\ }\href@noop
  {} {\bibinfo {title} {A solution to 2-copy distillability of {Werner}
  states}} (\bibinfo {year} {2026}),\ \Eprint
  {https://arxiv.org/abs/2607.21367} {arXiv:2607.21367} \BibitemShut {NoStop}%
\bibitem [{\citenamefont {Song}\ and\ \citenamefont {Chen}(2026)}]{sc}%
  \BibitemOpen
  \bibfield  {author} {\bibinfo {author} {\bibfnamefont {Z.}~\bibnamefont
  {Song}}\ and\ \bibinfo {author} {\bibfnamefont {L.}~\bibnamefont {Chen}},\
  }\href@noop {} {\bibinfo {title} {A partial-trace matrix inequality and
  {Werner}-state distillability}} (\bibinfo {year} {2026}),\ \Eprint
  {https://arxiv.org/abs/2607.23416} {arXiv:2607.23416} \BibitemShut {NoStop}%
\bibitem [{\citenamefont {Fraser}\ \emph {et~al.}(2026)\citenamefont {Fraser},
  \citenamefont {Huber}, \citenamefont {Pozsgay},\ and\ \citenamefont
  {Vona}}]{fraser}%
  \BibitemOpen
  \bibfield  {author} {\bibinfo {author} {\bibfnamefont {T.~C.}\ \bibnamefont
  {Fraser}}, \bibinfo {author} {\bibfnamefont {F.}~\bibnamefont {Huber}},
  \bibinfo {author} {\bibfnamefont {B.}~\bibnamefont {Pozsgay}},\ and\ \bibinfo
  {author} {\bibfnamefont {I.}~\bibnamefont {Vona}},\ }\href@noop {} {\bibinfo
  {title} {On the two-copy distillability of {Werner} states and a new partial
  trace inequality}} (\bibinfo {year} {2026}),\ \Eprint
  {https://arxiv.org/abs/2607.24309} {arXiv:2607.24309} \BibitemShut {NoStop}%
\bibitem [{\citenamefont {Bharti}\ \emph {et~al.}(2026)\citenamefont {Bharti},
  \citenamefont {Gajjala},\ and\ \citenamefont {Haug}}]{bharti}%
  \BibitemOpen
  \bibfield  {author} {\bibinfo {author} {\bibfnamefont {K.}~\bibnamefont
  {Bharti}}, \bibinfo {author} {\bibfnamefont {R.}~\bibnamefont {Gajjala}},\
  and\ \bibinfo {author} {\bibfnamefont {T.}~\bibnamefont {Haug}},\ }\href@noop
  {} {\bibinfo {title} {Two-copy nondistillability of {Werner} states: Sharp
  partial-trace inequalities and finite-copy extensions}} (\bibinfo {year}
  {2026}),\ \Eprint {https://arxiv.org/abs/2607.24479} {arXiv:2607.24479}
  \BibitemShut {NoStop}%
\bibitem [{\citenamefont {MacCormack}\ and\ \citenamefont
  {Bao}(2026)}]{maccormackbao}%
  \BibitemOpen
  \bibfield  {author} {\bibinfo {author} {\bibfnamefont {I.}~\bibnamefont
  {MacCormack}}\ and\ \bibinfo {author} {\bibfnamefont {N.}~\bibnamefont
  {Bao}},\ }\href
  {https://github.com/physical-superintelligence/oqp-batch-2026/blob/7049357c42a8fe85262da08fdc10d57b4d9e67a0/oqp02-three-copy-werner/paper/oqp02-three-copy-werner.pdf}
  {\bibinfo {title} {Three copies of the {Werner} state: The sharp curve in
  dimensions three and four and a bound in every dimension}} (\bibinfo {year}
  {2026}),\ \bibinfo {note} {preprint, 23 October 2026}\BibitemShut {NoStop}%
\bibitem [{\citenamefont {Tomar}(2026{\natexlab{a}})}]{tomar}%
  \BibitemOpen
  \bibfield  {author} {\bibinfo {author} {\bibfnamefont {T.~S.}\ \bibnamefont
  {Tomar}},\ }\href {https://hexagonmath.org/2611.00464v1} {\bibinfo {title}
  {Undistillability of {Werner} states from three, four and five copies}}
  (\bibinfo {year} {2026}{\natexlab{a}}),\ \bibinfo {note} {{Hexagon} preprint
  2611.00464v1, November 15, 2026}\BibitemShut {NoStop}%
\bibitem [{\citenamefont {Tomar}(2026{\natexlab{b}})}]{tomarendpoint}%
  \BibitemOpen
  \bibfield  {author} {\bibinfo {author} {\bibfnamefont {T.~S.}\ \bibnamefont
  {Tomar}},\ }\href {https://hexagonmath.org/2611.00466v1} {\bibinfo {title}
  {Equality cases and every-copy results for the {Werner} endpoint inequality}}
  (\bibinfo {year} {2026}{\natexlab{b}}),\ \bibinfo {note} {{Hexagon} preprint
  2611.00466v1, November 15, 2026}\BibitemShut {NoStop}%
\bibitem [{\citenamefont {Chitambar}\ \emph {et~al.}(2014)\citenamefont
  {Chitambar}, \citenamefont {Leung}, \citenamefont {Man{\v{c}}inska},
  \citenamefont {Ozols},\ and\ \citenamefont {Winter}}]{locc}%
  \BibitemOpen
  \bibfield  {author} {\bibinfo {author} {\bibfnamefont {E.}~\bibnamefont
  {Chitambar}}, \bibinfo {author} {\bibfnamefont {D.}~\bibnamefont {Leung}},
  \bibinfo {author} {\bibfnamefont {L.}~\bibnamefont {Man{\v{c}}inska}},
  \bibinfo {author} {\bibfnamefont {M.}~\bibnamefont {Ozols}},\ and\ \bibinfo
  {author} {\bibfnamefont {A.}~\bibnamefont {Winter}},\ }\bibfield  {title}
  {\bibinfo {title} {Everything you always wanted to know about {LOCC} (but
  were afraid to ask)},\ }\href {https://doi.org/10.1007/s00220-014-1953-9}
  {\bibfield  {journal} {\bibinfo  {journal} {Commun. Math. Phys.}\ }\textbf
  {\bibinfo {volume} {328}},\ \bibinfo {pages} {303} (\bibinfo {year}
  {2014})}\BibitemShut {NoStop}%
\bibitem [{\citenamefont {Eggeling}\ \emph {et~al.}(2001)\citenamefont
  {Eggeling}, \citenamefont {Vollbrecht}, \citenamefont {Werner},\ and\
  \citenamefont {Wolf}}]{eggeling}%
  \BibitemOpen
  \bibfield  {author} {\bibinfo {author} {\bibfnamefont {T.}~\bibnamefont
  {Eggeling}}, \bibinfo {author} {\bibfnamefont {K.~G.~H.}\ \bibnamefont
  {Vollbrecht}}, \bibinfo {author} {\bibfnamefont {R.~F.}\ \bibnamefont
  {Werner}},\ and\ \bibinfo {author} {\bibfnamefont {M.~M.}\ \bibnamefont
  {Wolf}},\ }\bibfield  {title} {\bibinfo {title} {Distillability via protocols
  respecting the positivity of partial transpose},\ }\href
  {https://doi.org/10.1103/PhysRevLett.87.257902} {\bibfield  {journal}
  {\bibinfo  {journal} {Phys. Rev. Lett.}\ }\textbf {\bibinfo {volume} {87}},\
  \bibinfo {pages} {257902} (\bibinfo {year} {2001})}\BibitemShut {NoStop}%
\bibitem [{\citenamefont {Vollbrecht}\ and\ \citenamefont
  {Wolf}(2002)}]{vollbrecht}%
  \BibitemOpen
  \bibfield  {author} {\bibinfo {author} {\bibfnamefont {K.~G.~H.}\
  \bibnamefont {Vollbrecht}}\ and\ \bibinfo {author} {\bibfnamefont {M.~M.}\
  \bibnamefont {Wolf}},\ }\bibfield  {title} {\bibinfo {title} {Activating
  distillation with an infinitesimal amount of bound entanglement},\ }\href
  {https://doi.org/10.1103/PhysRevLett.88.247901} {\bibfield  {journal}
  {\bibinfo  {journal} {Phys. Rev. Lett.}\ }\textbf {\bibinfo {volume} {88}},\
  \bibinfo {pages} {247901} (\bibinfo {year} {2002})}\BibitemShut {NoStop}%
\bibitem [{\citenamefont {Shor}\ \emph {et~al.}(2001)\citenamefont {Shor},
  \citenamefont {Smolin},\ and\ \citenamefont {Terhal}}]{sst}%
  \BibitemOpen
  \bibfield  {author} {\bibinfo {author} {\bibfnamefont {P.~W.}\ \bibnamefont
  {Shor}}, \bibinfo {author} {\bibfnamefont {J.~A.}\ \bibnamefont {Smolin}},\
  and\ \bibinfo {author} {\bibfnamefont {B.~M.}\ \bibnamefont {Terhal}},\
  }\bibfield  {title} {\bibinfo {title} {Nonadditivity of bipartite distillable
  entanglement follows from a conjecture on bound entangled {Werner} states},\
  }\href {https://doi.org/10.1103/PhysRevLett.86.2681} {\bibfield  {journal}
  {\bibinfo  {journal} {Phys. Rev. Lett.}\ }\textbf {\bibinfo {volume} {86}},\
  \bibinfo {pages} {2681} (\bibinfo {year} {2001})}\BibitemShut {NoStop}%
\bibitem [{\citenamefont {Watrous}(2018)}]{watrous}%
  \BibitemOpen
  \bibfield  {author} {\bibinfo {author} {\bibfnamefont {J.}~\bibnamefont
  {Watrous}},\ }\href@noop {} {\emph {\bibinfo {title} {The Theory of Quantum
  Information}}}\ (\bibinfo  {publisher} {Cambridge University Press},\
  \bibinfo {address} {Cambridge},\ \bibinfo {year} {2018})\BibitemShut
  {NoStop}%
\bibitem [{\citenamefont {Horodecki}\ \emph {et~al.}(1997)\citenamefont
  {Horodecki}, \citenamefont {Horodecki},\ and\ \citenamefont
  {Horodecki}}]{horodecki97}%
  \BibitemOpen
  \bibfield  {author} {\bibinfo {author} {\bibfnamefont {M.}~\bibnamefont
  {Horodecki}}, \bibinfo {author} {\bibfnamefont {P.}~\bibnamefont
  {Horodecki}},\ and\ \bibinfo {author} {\bibfnamefont {R.}~\bibnamefont
  {Horodecki}},\ }\bibfield  {title} {\bibinfo {title} {Inseparable two
  spin-$\frac12$ density matrices can be distilled to a singlet form},\ }\href
  {https://doi.org/10.1103/PhysRevLett.78.574} {\bibfield  {journal} {\bibinfo
  {journal} {Phys. Rev. Lett.}\ }\textbf {\bibinfo {volume} {78}},\ \bibinfo
  {pages} {574} (\bibinfo {year} {1997})}\BibitemShut {NoStop}%
\bibitem [{\citenamefont {Bennett}\ \emph {et~al.}(1996)\citenamefont
  {Bennett}, \citenamefont {DiVincenzo}, \citenamefont {Smolin},\ and\
  \citenamefont {Wootters}}]{bdsw}%
  \BibitemOpen
  \bibfield  {author} {\bibinfo {author} {\bibfnamefont {C.~H.}\ \bibnamefont
  {Bennett}}, \bibinfo {author} {\bibfnamefont {D.~P.}\ \bibnamefont
  {DiVincenzo}}, \bibinfo {author} {\bibfnamefont {J.~A.}\ \bibnamefont
  {Smolin}},\ and\ \bibinfo {author} {\bibfnamefont {W.~K.}\ \bibnamefont
  {Wootters}},\ }\bibfield  {title} {\bibinfo {title} {Mixed-state entanglement
  and quantum error correction},\ }\href
  {https://doi.org/10.1103/PhysRevA.54.3824} {\bibfield  {journal} {\bibinfo
  {journal} {Phys. Rev. A}\ }\textbf {\bibinfo {volume} {54}},\ \bibinfo
  {pages} {3824} (\bibinfo {year} {1996})}\BibitemShut {NoStop}%
\bibitem [{\citenamefont {Gatermann}\ and\ \citenamefont
  {Parrilo}(2004)}]{gatermann}%
  \BibitemOpen
  \bibfield  {author} {\bibinfo {author} {\bibfnamefont {K.}~\bibnamefont
  {Gatermann}}\ and\ \bibinfo {author} {\bibfnamefont {P.~A.}\ \bibnamefont
  {Parrilo}},\ }\bibfield  {title} {\bibinfo {title} {Symmetry groups,
  semidefinite programs, and sums of squares},\ }\href
  {https://doi.org/10.1016/j.jpaa.2003.12.011} {\bibfield  {journal} {\bibinfo
  {journal} {J. Pure Appl. Algebra}\ }\textbf {\bibinfo {volume} {192}},\
  \bibinfo {pages} {95} (\bibinfo {year} {2004})}\BibitemShut {NoStop}%
\bibitem [{\citenamefont {Bareiss}(1968)}]{bareiss}%
  \BibitemOpen
  \bibfield  {author} {\bibinfo {author} {\bibfnamefont {E.~H.}\ \bibnamefont
  {Bareiss}},\ }\bibfield  {title} {\bibinfo {title} {Sylvester's identity and
  multistep integer-preserving {G}aussian elimination},\ }\href
  {https://doi.org/10.1090/S0025-5718-1968-0226829-0} {\bibfield  {journal}
  {\bibinfo  {journal} {Math. Comp.}\ }\textbf {\bibinfo {volume} {22}},\
  \bibinfo {pages} {565} (\bibinfo {year} {1968})}\BibitemShut {NoStop}%
\bibitem [{\citenamefont {Fulton}\ and\ \citenamefont
  {Harris}(1991)}]{fultonharris}%
  \BibitemOpen
  \bibfield  {author} {\bibinfo {author} {\bibfnamefont {W.}~\bibnamefont
  {Fulton}}\ and\ \bibinfo {author} {\bibfnamefont {J.}~\bibnamefont
  {Harris}},\ }\href@noop {} {\emph {\bibinfo {title} {Representation Theory: A
  First Course}}}\ (\bibinfo  {publisher} {Springer},\ \bibinfo {address} {New
  York},\ \bibinfo {year} {1991})\BibitemShut {NoStop}%
\bibitem [{\citenamefont {Gurvits}\ and\ \citenamefont
  {Barnum}(2002)}]{gurvits}%
  \BibitemOpen
  \bibfield  {author} {\bibinfo {author} {\bibfnamefont {L.}~\bibnamefont
  {Gurvits}}\ and\ \bibinfo {author} {\bibfnamefont {H.}~\bibnamefont
  {Barnum}},\ }\bibfield  {title} {\bibinfo {title} {Largest separable balls
  around the maximally mixed bipartite quantum state},\ }\href
  {https://doi.org/10.1103/PhysRevA.66.062311} {\bibfield  {journal} {\bibinfo
  {journal} {Phys. Rev. A}\ }\textbf {\bibinfo {volume} {66}},\ \bibinfo
  {pages} {062311} (\bibinfo {year} {2002})},\ \bibinfo {note}
  {arXiv:quant-ph/0204159}\BibitemShut {NoStop}%
\bibitem [{\citenamefont {M{\"u}ller-Hermes}\ \emph {et~al.}(2016)\citenamefont
  {M{\"u}ller-Hermes}, \citenamefont {Reeb},\ and\ \citenamefont {Wolf}}]{mrw}%
  \BibitemOpen
  \bibfield  {author} {\bibinfo {author} {\bibfnamefont {A.}~\bibnamefont
  {M{\"u}ller-Hermes}}, \bibinfo {author} {\bibfnamefont {D.}~\bibnamefont
  {Reeb}},\ and\ \bibinfo {author} {\bibfnamefont {M.~M.}\ \bibnamefont
  {Wolf}},\ }\bibfield  {title} {\bibinfo {title} {Positivity of linear maps
  under tensor powers},\ }\href {https://doi.org/10.1063/1.4927070} {\bibfield
  {journal} {\bibinfo  {journal} {J. Math. Phys.}\ }\textbf {\bibinfo {volume}
  {57}},\ \bibinfo {pages} {015202} (\bibinfo {year} {2016})}\BibitemShut
  {NoStop}%
\end{thebibliography}%

\end{document}